\documentclass[twocolumn, twocolappendix]{aastex63}
\let\tablenum\relax

\received{XXX}
\revised{YYY}
\accepted{ZZZ}

\submitjournal{ApJ}

\shorttitle{LFs and clustering of HAEs}
\shortauthors{Lin et al.}

\graphicspath{{./}{figures/}}
 
\usepackage{amsmath,bm}
\usepackage{multirow}
\usepackage{hyperref}
\usepackage{nameref}
\usepackage{siunitx}

\newcommand{\ha}{H$\alpha$}
\newcommand{\chisq}{$\chi^2$}

\newcommand{\Nemi}{{3000}}
\newcommand{\NHAEtot}{{936}}
\newcommand{\NHAE}{{888}}
\newcommand{\NHAEcl}{{719}}

\newcommand{\xj}[1]{{{#1}}}

\begin{document}

\title{The Luminosity Function and Clustering of H$\alpha$ Emitting Galaxies at $z\approx4-6$ \\ from a Complete NIRCam Grism Redshift Survey}

\author[0000-0001-6052-4234]{Xiaojing Lin}
\affiliation{Department of Astronomy, Tsinghua University, Beijing 100084, China}
\email{xiaojinglin.astro@gmail.com}
\affil{Steward Observatory, University of Arizona, 933 N Cherry Ave, Tucson, AZ 85721, USA}

\author[0000-0003-1344-9475]{Eiichi Egami}
\affiliation{Steward Observatory, University of Arizona, 933 N Cherry Ave, Tucson, AZ 85721, USA}

\author[0000-0002-4622-6617]{Fengwu Sun}
\affiliation{Center for Astrophysics $|$ Harvard \& Smithsonian, 60 Garden St., Cambridge, MA 02138, USA}

\author[0000-0002-4321-3538]{Haowen Zhang}
\affiliation{Steward Observatory, University of Arizona, 933 N Cherry Ave, Tucson, AZ 85721, USA}

\author[0000-0003-3310-0131]{Xiaohui Fan}
\affiliation{Steward Observatory, University of Arizona, 933 N Cherry Ave, Tucson, AZ 85721, USA}

\author[0000-0003-4337-6211]{Jakob M.\ Helton}
\affiliation{Steward Observatory, University of Arizona, 933 N Cherry Ave, Tucson, AZ 85721, USA}

\author[0000-0002-7633-431X]{Feige Wang}
\affiliation{Department of Astronomy, University of Michigan, 1085 South University Avenue, Ann Arbor, MI 48109, USA}


\author[0000-0002-8651-9879]{Andrew J.\ Bunker }
\affiliation{Department of Physics, University of Oxford, Denys Wilkinson Building, Keble Road, Oxford OX13RH, UK}

\author[0000-0001-8467-6478]{Zheng Cai}
\affiliation{Department of Astronomy, Tsinghua University, Beijing 100084, China}

\author[0000-0002-2929-3121]{Daniel J.\ Eisenstein}
\affiliation{Center for Astrophysics $|$ Harvard \& Smithsonian, 60 Garden St., Cambridge, MA 02138, USA}

\author{Daniel T. Jaffe}
\affiliation{The University of Texas at Austin, Department of Astronomy, 2515 Speedway, Stop C1400, Austin, TX 78712-1205, USA}

\author[0000-0001-7673-2257]{Zhiyuan Ji}
\affiliation{Steward Observatory, University of Arizona, 933 N Cherry Ave, Tucson, AZ 85721, USA}

\author[0000-0002-5768-738X]{Xiangyu Jin}
\affiliation{Steward Observatory, University of Arizona, 933 N Cherry Ave, Tucson, AZ 85721, USA}

\author[0000-0003-4924-5941]{Maria Anne Pudoka}
\affiliation{Steward Observatory, University of Arizona, 933 N Cherry Ave, Tucson, AZ 85721, USA}

\author[0000-0002-8224-4505]{Sandro Tacchella}
\affiliation{Kavli Institute for Cosmology, University of Cambridge, Madingley Road, Cambridge, CB3 0HA, UK
}
\affiliation{Cavendish Laboratory, University of Cambridge, 19 JJ Thomson Avenue, Cambridge, CB3 0HE, UK}

\author[0000-0003-0747-1780]{Wei Leong Tee
}
\affiliation{Steward Observatory, University of Arizona, 933 N Cherry Ave, Tucson, AZ 85721, USA}

\author[0000-0002-5104-8245]{Pierluigi Rinaldi}
\affiliation{Steward Observatory, University of Arizona, 933 N Cherry Ave, Tucson, AZ 85721, USA}

\author[0000-0002-4271-0364]{Brant Robertson}
 \affiliation{Department of Astronomy and Astrophysics, University of California, Santa Cruz, 1156 High Street, Santa Cruz, CA 95064, USA}

\author[0000-0001-6561-9443]{Yang Sun}
\affiliation{Steward Observatory, University of Arizona, 933 N Cherry Ave, Tucson, AZ 85721, USA}

\author[0000-0001-9262-9997]{Christopher N.\ A.\ Willmer}
\affiliation{Steward Observatory, University of Arizona, 933 N Cherry Ave, Tucson, AZ 85721, USA}

\author[0000-0002-4201-7367]{Chris Willott}
\affiliation{NRC Herzberg, 5071 West Saanich Rd, Victoria, BC V9E 2E7, Canada}

\author[0000-0002-1574-2045]{Junyu Zhang}
\affiliation{Steward Observatory, University of Arizona, 933 N Cherry Ave, Tucson, AZ 85721, USA}

\author[0000-0003-3307-7525]{Yongda Zhu}
\affiliation{Steward Observatory, University of Arizona, 933 N Cherry Ave, Tucson, AZ 85721, USA}

\correspondingauthor{Xiaojing Lin}

\begin{abstract}
We study the luminosity function (LF) and clustering properties of \NHAE\ H$\alpha$-emitters (HAEs) at $3.75 < z < 6$ in the GOODS-N field. The sample, built from JWST CONGRESS and FRESCO NIRCam grism surveys using a novel redshift assignment algorithm, spans $\sim$62 arcmin$^2$ and reaches $L_{\rm H\alpha} \sim 10^{41.2}$\,\si{erg\,s^{-1}}. We identify two prominent filamentary protoclusters at $z \approx 4.41$ and $z \approx 5.19$, hosting 98 and 144 HAEs, respectively. The observed H$\alpha$ LFs have similarly shallow faint-end slopes ($\alpha \approx -1.2$ to $-1.3$) for both protocluster and field galaxies at $3.75 < z < 5$, and for protoclusters at $5 < z < 6$. In contrast, the field LF at $5 < z < 6$ is much steeper ($\alpha = -1.87_{-0.23}^{+0.30}$), indicating that protocluster galaxies at $z > 5$ are more evolved, resembling lower-$z$ populations.
The observed star formation rate density, integrated down to 0.45 \si{M_\odot\,yr^{-1}}, is $0.050^{+0.002}_{-0.003}\, M_\odot \, {\rm yr}^{-1} \, {\rm Mpc}^{-3}$ at $3.75 < z < 5$ and $0.046^{+0.006}_{-0.004}\, M_\odot \, {\rm yr}^{-1} \, {\rm Mpc}^{-3}$ at $5 < z < 6$,  with protoclusters contributing about 25\% and 55\%, respectively. We conduct the star-formation-rate-limited three-dimensional clustering analysis at $z > 4$. The filamentary geometry of protoclusters flattens the power-law shape of the HAE auto-correlation functions. The correlation function of field HAEs has correlation length of $r_0 = 4.61^{+1.00}_{-0.68}\,h^{-1}{\rm Mpc}$ at $z \approx 4-5$ and $r_0 = 6.23^{+1.68}_{-1.13}\,h^{-1}{\rm Mpc}$ at $z \approx 5-6$. Comparing with the \textsc{UniverseMachine} simulation, we infer the dark matter (sub-)halo masses of HAEs to be $\log (M_h/M_\odot)=11.0-11.2$ at $z\approx 4-6$, with a scatter of 0.4 dex.
\end{abstract}


\section{Introduction}

The \ha\ emission line in galaxies has been well established as a reliable indicator of their global star formation rates (SFR) \citep{Kennicutt1983, Gallego1995, Kennicutt1998}. Originating from hydrogen recombination triggered by ionizing photons from young massive stars, \ha\ is sensitive to star formation within the past 10 Myr \citep{Murphy2011, Kennicutt2012, Haydon2020, FloresVelazquez2021}. Unlike the UV continuum, which traces star formation over a longer timescale of $<$100 Myr, \ha\ is less affected by dust attenuation. These characteristics make \ha\ one of the most powerful tracers of cosmic star-formation activity.

Over the past two decades, numerous wide-field surveys have mapped the cosmic star formation density up to $z\sim3$ through \ha\, using either narrow-band imaging or spectroscopic observations \citep[e.g.,][]{Ly2007, Geach2008, Hayes2010, Lee2012, Sobral2013, Pirzkal2013, Stroe2015, Sobral2016, Nagaraj2023}. These studies have revealed the evolution of \ha\ luminosity functions (LFs) and the buildup of star formation history until its peak at $z\sim1-3$ \citep{Madau2014}.  In addition to the redshift evolution of star formation density, the impact of large-scale environments on shaping the global star formation has also been extensively studied \citep[e.g.,][]{Sobral2011, Koyama2013, Stroe2017, Shimakawa2018}.  Clustering analysis of \ha\ emitting galaxies, such as their auto-correlation function,  provides constraints on their host dark matter halo masses and places star formation within the framework of the $\Lambda$CDM cosmology \citep{Sobral2010, Geach2012, Coil2017, Cochrane2018, Berti2019}. At $z \lesssim 2.3$, the clustering of H$\alpha$ emitters exhibits a strong dependence on their H$\alpha$ luminosities, indicating that more star formation occurs in more massive halos \citep{Sobral2010, Stroe2015, Cochrane2017, Cochrane2018, Clontz2022}. These findings highlight the role of dark matter halos in shaping the history of cosmic star formation and structure formation.

At $z\gtrsim3$, the \ha\ lines are redshifted beyond the scope of ground-based facilities and HST grism spectroscopy \citep[e.g.,][]{Atek2010, Momcheva2016}. Before the launch of JWST, Spitzer/IRAC broadband photometry at 3.6 $\mu$m and 4.5 $\mu$m provided the only opportunity to probe \ha\ emission at high redshifts \citep[e.g.,][]{Schaerer2010, Shim2011, Rasappu2016, Stefanon2022, Bollo2023}. The inferred \ha\ luminosity through IRAC color excess is highly uncertain, especially in the absence of knowledge about the stellar populations and, consequently, the spectral shapes of these high-redshift galaxies.  Although one can alternatively use UV luminosity to trace the cosmic star formation history, it is highly sensitive to dust attenuation, which remains poorly understood in the early Universe \citep[e.g.,][]{Overzier2011, Buat2012, Dominguez2013, Castellano2014, Reddy2015}. 

On the other hand, the lack of precise redshift for these \ha\ emitters at  $z>3$ makes it impossible to conduct SFR-limited three-dimensional clustering analyses. Studies on the clustering of \ha\ emitters have been limited to $z \approx 2.23$ and have not extended beyond this redshift for more than ten years \citep{Geach2012}. During these ten years, Ly$\alpha$ emitters have become the most widely used tracers for dark matter halos at $z > 3$ with spectroscopic redshifts available \citep{Diener2017, HerreroAlonso2023}. However, they are biased tracers of star formation due to their degeneracy with radiative transfer and escape processes from the interstellar medium \citep{Zheng2010, Dijkstra2017}. These complex physical processes, which introduce non-linear effects, make it challenging to interpret the clustering of Ly$\alpha$ emitters in the context of dark matter halos and cosmology \citep[e.g.,][]{Zheng2011, Hutter2015, Behrens2018, Byrohl2019, Im2024}. 

The advent of JWST \citep{Gardner2023} ushers in a new era for \ha\ studies in the high-redshift Universe \citep[e.g.,][]{Rinaldi2023, Cameron2023, Bunker2024, Simmonds2024, JADES_DR3, Fu2025}. The unprecedented infrared capabilities of JWST have revealed thousands of $z>3$ \ha\ emission lines with NIRCam median-band imaging \citep[e.g.,][]{Williams2023, Simmonds2024}, NIRSpec, NIRISS, and NIRCam/Wide Field Slitless Spectroscopy (WFSS). In particular, the NIRCam/WFSS efficiently captures the emission lines of all galaxies within the field of view, producing a blindly-searched emitter sample with a simple flux-limited selection function \citep{Greene2017, Rieke2023}. Since JWST Cycle-1, numerous NIRCam/WFSS programs have demonstrated their power in assembling large samples of \ha\ emitters \citep[e.g.,][]{Tang2024, Lin2024a, Naidu2024}. These grism observations enable precise measurement of \ha\ LFs by directly detecting the \ha\ lines \citep{Sun2023, CoveloPaz2024, Fu2025}, thereby better quantifying the cosmic star formation density. They also directly map large-scale structures \citep{Helton2024a, Helton2024, Sun2024, Herard_Demanche_2023}, enabling galaxy clustering analyses with accurate spectroscopic redshifts \citep{Eilers2024, Matthee2024b, Shuntov2025}. 

In this study, we analyze over 800 \ha\ emitters identified from the Complete NIRCam Grism Redshift Survey (CONGRESS) in the GOODS-N field. The dataset includes grism observations from the Cycle-1 program FRESCO \citep[``First Reionization Epoch Spectroscopically Complete Observations", GO-1895, PI Oesch,][]{Oesch2023} and the Cycle-2 program CONGRESS (GO-3577, PI Eiichi \& Sun; Sun et al. in prep). FRESCO covers 62 arcmin$^2$ of the GOODS-N field with the F444W grism, and CONGRESS targets the same region with the F356W grism. Together, both programs provide a continuous wavelength coverage of $3.1$–$5.0$\,\micron. The F356W grism of CONGRESS not only expands the sample but also captures the [\ion{O}{3}] doublet of $z>5$ \ha\ emitters. These [\ion{O}{3}] lines typically have higher fluxes than \ha\ and, when combined with \ha, provide high-fidelity redshift measurements through multiple emission lines. In this paper, we aim to precisely measure the \ha\ LFs at $z\approx4-6$, tracing the evolution of cosmic star formation before its peak at $z\sim2$. We then investigate the clustering properties of \ha\ emitters at $z \approx 4-6$, extending the SFR-limited three-dimensional clustering studies to $z \gtrsim 4$ for the first time.

This paper outline is as follows. In \S\ref{sec:datasample}, we describe the data used and the selection of \ha\ emitters. In \S\ref{sec:HaLF}, we present the \ha\ LFs and the constraints on the star formation history. In \S\ref{sec:clustering}, we compute the three-dimensional two-point auto-correlation function, analyze the clustering properties, and discuss the constraints on the dark matter halo masses. We provide the technical details in the Appendix, including the emitter selection algorithm, the completeness correction method, and other relevant procedures. Throughout this work, a flat $\Lambda$CDM cosmology is assumed, with $\rm H_0 = 70~km~s^{-1}~Mpc^{-1} $, $\Omega_{\Lambda,0} = 0.7$ and  $\Omega_{m,0}=0.3$.  We define $h={\rm H_0} / 100 = 0.7$.

\bigskip

\section{Data and \ha\ Sample}\label{sec:datasample}

\subsection{Imaging and photometric catalog}
We use the JWST images and photometric catalog in the GOODS-N field as part of the JADES Data Release 3\footnote{ available at \url{https://archive.stsci.edu/hlsp/jades}} \citep[DR3;][]{JADES_DR3}.  The JADES JWST/NIRCam images in the GOODS-N field include observations from GTO program 1181 (PI Eisenstein) and GO program 1895 (FRESCO, PI Oesch), spanning F090W, F115W, F150W, F182M, F200W, F210M, F277W, F335M, F356W, F410M, and F444W. The final mosaics cover an area of 47-63 arcmin$^2$ from F090W to F410M, and 83 arcmin$^2$ in F444W \citep{JADES_DR3}. We refer to \cite{JADES_Eisenstein2023} for a detailed description of the JADES survey design, \cite{Rieke2023} and Robertson et al. (in prep) for the imaging data reduction, and \cite{JADES_DR3} for the content of DR3. For the direct images used for the grism spectra, the JADES images reach a median 5$\sigma$ point-source depth of 29.38 mag in F444W and 29.97 mag in F356W, with aperture-corrected photometry using an $r=0.15\arcsec$ circular aperture \xj{(see Table 1 of \citealt{JADES_DR3}).}

The JADES GOODS-N photometric catalog includes multi-band photometry of 11 JWST/NIRCam filters, as mentioned above, and five HST/ACS filters (F435W, F606W, F775W, F814W, and F850LP). The HST/ACS photometry is based on images from the Hubble Legacy Fields project \citep[HLF,][]{Illingworth2017}. We refer to \cite{Robertson2024} for detailed source detection and photometry measurement methods.  \xj{The photometric redshifts are derived using \texttt{EAZY} \citep{Brammer2008}, based on circular aperture photometry with a radius of $r = 0\farcs1$. The aperture corrections have been applied using the point spread functions (PSFs) corresponding to each band. The spectral energy distribution (SED) templates used in the fitting are optimized for high-redshift galaxy populations, as described in \citet{Hainline2024}. This template set includes the original \texttt{EAZY v1.3} templates, supplemented by seven additional templates generated with the Flexible Stellar Population Synthesis \citep[\texttt{fsps},][]{Conroy2010}, specifically designed to improve photometric redshift estimates for \textsc{jaguar} mock galaxies \citep{Williams2018}.}

\subsection{JWST/NIRCam grism spectroscopy}
The JWST/NIRCam WFSS observations were conducted in the GOODS-N field in both the F356W and F444W filters. The Cycle-1 program, FRESCO, covers 62 arcmin$^{2}$ of the GOODS-N field through the F444W row-direction grism (Grism R). The FRESCO observations include eight pointings, each with an exposure time of 8$\times$880\,s. The Cycle-2 program, CONGRESS, intentionally targets the same areas in the  GOODS-N field observed by FRESCO (but with a 2-degree offset in positional angle) using the F356W Grism R. It includes 12 pointings, and the exposure time is 8$\times$472\,s per pointing.   Combining the grism observations in F356W and F444W  leads to a total wavelength range of 3.1--5.0\,\micron. The average $5\sigma$ emission line flux limit for both the F356W and F444W grism is about $2\times 10^{-18}$ \si{erg\,s^{-1}\,cm^{-2}}. The overlap area between JADES DR3 images and FRESCO+CONGRESS WFSS is about 62 arcmin$^2$, as illustrated in the left panel of Figure \ref{fig:distribution}.

The grism data reduction and spectral extraction used in this study are described in Sun et al. (prep), following the standard routine outlined in \citet{Sun2023}. Here we briefly summarize the procedure below. For individual exposures of grism data and their corresponding short-wavelength (SW) direct images, we perform flat-fielding, subtract the sigma-clipped median sky background, and align the WCS frames. We then measure the astrometric offsets between the SW direct images and the JADES DR3 GOODS-N catalog. The offsets are added to the spectral tracing model for accurate wavelength calibrations.  The spectral tracing and wavelength calibration are based on Commissioning, Cycle-1, and Cycle-2 calibration data taken in the SMP-LMC-58 field in June 2024 (PID 1076, 1479, 1480 and 4449; see \citealt{sapphires_edr} for the most recent update)\footnote{\url{https://github.com/fengwusun/nircam_grism}}.
The flux calibration is based on Cycle-1 calibration data (PID: 1076, 1536, 1537, 1538).   We optimally extract the 1D spectra based on the source morphology \citep{Horne1986}. To better remove contamination from other sources along the dispersion direction, we extract three versions of the spectra: (1) \textsc{mod-A-only} which includes only the spectra extracted from NIRCam module A, and (2) \textsc{mod-B-only} which includes only the spectra extracted from NIRCam module B. (3) The final \textsc{coadd} spectra combine all available exposures. For sources located in the overlapping regions of both modules from different pointings, the \textsc{coadd} spectra incorporate data from both \textsc{mod-A-only} and \textsc{mod-B-only}.

\subsection{Selection of emission line galaxies}\label{sec:emission_line_selection}

We develop a semi-automated algorithm to identify line emitters and determine their redshifts. The algorithm automatically assigns the most probable redshifts based on emission lines in the grism spectra. It greatly reduces the workload for visual inspection, though manual verification remains a necessary final step. Detailed procedures of this algorithm are presented in Appendix \ref{sec:redshift_algorithm}. A brief summary is provided below.

For emission line detection, we perform median filtering on the 2D spectra along the dispersion direction and extract the continuum-removed 1D spectra.  We search for emission lines on the continuum-removed 1D spectra by detecting high signal-to-noise ratio (S/N$>5$) flux peaks and on the continuum-removed 2D spectra by running source detection. The 1D emission lines are modeled using a single Gaussian profile centered at the peak wavelength.  The line detection algorithm is detailed in \cite{Wang2023}.  The 1D and 2D line searches are conducted in \textsc{coadd}, \textsc{mod-A-only}, and \textsc{mod-B-only} spectra. The lines with fluxes brighter than the corresponding broadband photometry (F356W or F444W) are labeled as contamination and thus excluded. We further define an emission line as robust if it is simultaneously detected in the 1D and 2D spectra of \textsc{coadd}, and if available, in both the \textsc{mod-A-only} and \textsc{mod-B-only} extractions. The full criteria of robust lines are listed in Appendix \ref{sec:effective_line_table}.  We finally obtain a table of robust emission lines for each source, which contains the line wavelengths, fluxes, S/N, and FWHMs.

To determine the redshift solution for each emitter, we apply a two-step cross-correlation algorithm between the robust emission lines and a series of line templates for each source. We refer to Appendix \ref{sec:twostep_correlation} for a full description. The first step is a `position-only' correlation.  We cross-correlate the wavelengths of the detected robust emission lines with the template line wavelengths at $z=0.1-10$ with a step of $\delta z=0.1$. The results peak at several redshifts where the wavelengths of one or more detected lines match those of template lines. These peaks correspond to possible redshift solutions.  In the second step, we set finer redshift grids with $\delta z=0.001$ around the peaked $z$. We generate a model line spectrum based on the robust line table. We then construct a series of line bases, each with fixed line flux ratios determined empirically (e.g., H$\beta$+[\ion{O}{3}]$\lambda\lambda4960,5008$; \ha+[\ion{N}{2}]$\lambda\lambda6549,6585$+[\ion{S}{2}]$\lambda\lambda 6718,6732$). At each $z$ in the finer grid, we fit the line basis to the line spectrum model and calculate the \chisq. We finally obtain a series of \chisq\  as a function of $z$. We then select the best $z$ solution and score the confidence level based on their offset from the photo-$z$. The criteria for the confidence level is described in length in Appendix \ref{sec:score_system}.

We run the redshift determination algorithm for several iterations to finalize the redshift catalog. We first visually inspect sources with spectroscopic redshifts from the literature to verify the accuracy of the spec-$z$ derived from cross-correlation. The literature spec-$z$ is compiled from \cite{Wirth2004, Reddy2006, Barger2008, Cooper2011, Kriek2015, Barger2022}, along with the NIRSpec catalog from JADES DR3 \citep{JADES_DR3}. For sources with literature spec-$z$ from ground-based spectroscopy, we confirm their robustness if multiple lines are detected in the grism spectra. For sources with NIRSpec spec-$z$, we adopt the values determined from multiple lines in the prism spectra. This yields the first catalog of robust spec-$z$-confirmed emitters, including 281 sources from NIRSpec and $\sim600$ from ground-based spectra. We label the position of emission lines on the detector plane based on the grism tracing model and mask them if they appear in the spectra of other sources.  We run the redshift determination algorithm among the remaining sources,  select the emitters with the most confident redshift solutions, and visually inspect the robustness. We then label and mask the spec-$z$ confirmed emission lines, and start a new iteration.  After five iterations, we successfully identify over \Nemi\ emitters and robustly determine their redshift across $z=0-9$. The emitter catalog will be publicly released along with Sun et al.\ (in prep). The algorithm and code have also been applied to the Cycle 3 grism program SAPPHIRES \citep{sapphires_edr} and will be made publicly available upon the paper acceptance. Parts of the code are also incorporated into \texttt{unfold\_jwst} (Wang et al. in prep), a Python package that has been used to reduce grism data from multiple programs such as ASPIRE (GO-2078; \citealt{Wang2023}), NEXUS (GO-5105; \citealt{Shen2024}), and COSMOS-3D (GO-5893).

\bigskip

\section{HAEs, luminosity functions, and star formation rate density}\label{sec:HaLF}

\subsection{\ha\ emitting galaxies at $3.75<z<6$ }\label{sec:hae_sample}

\begin{figure*}[htbp!]
    \centering
    \includegraphics[width=\linewidth]{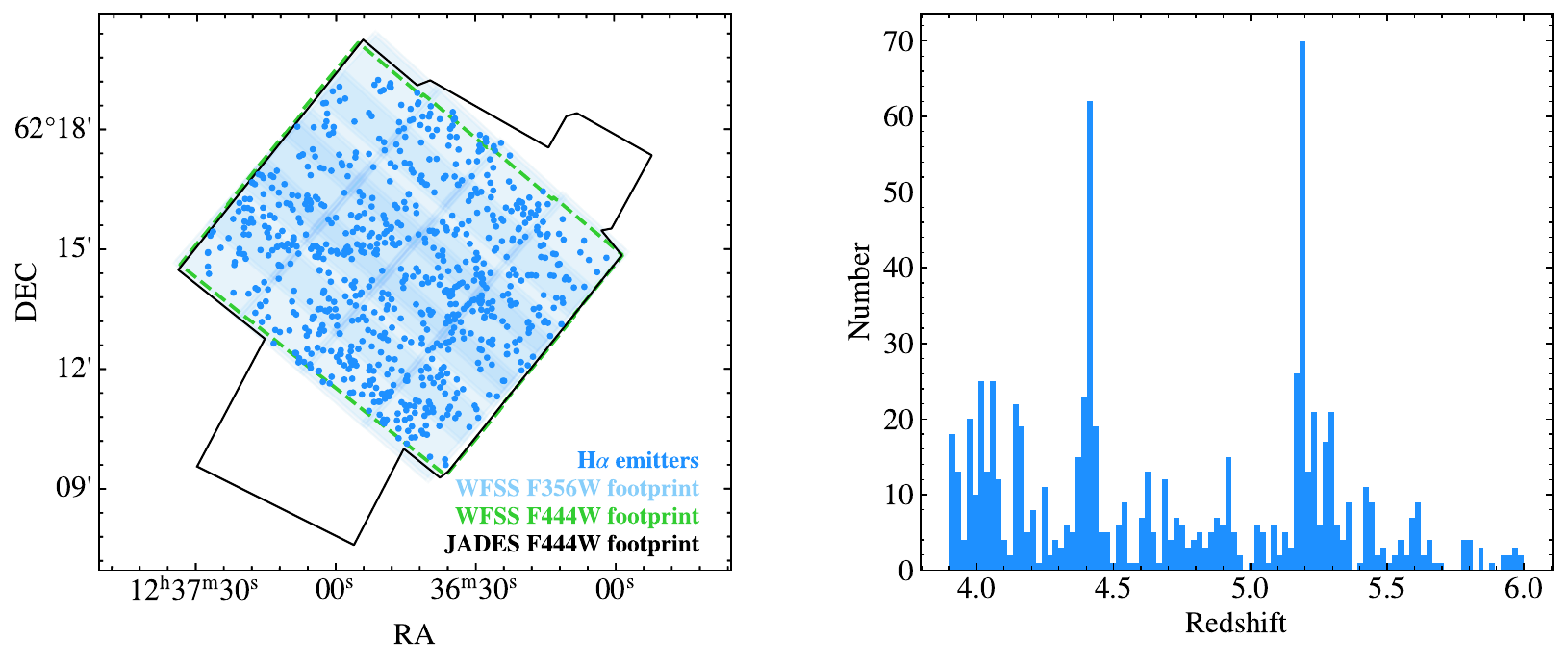}
    \caption{The spatial and redshift distributions of 888 HAEs at $3.75<z<6$ in the GOODS-N field. \textit{Left}: The distribution of HAEs within the JWST/NIRCam imaging and WFSS footprints. The blue-shaded region represents the F356W WFSS (GO-3577, CONGRESS) coverage \xj{and the green outline represents the F444W WFSS (GO-1895, FRESCO) coverage}.  The black outline shows the footprint of the F444W mosaic of JADES DR3 \citep{JADES_DR3}.  Blue dots represent HAEs spectroscopically confirmed from WFSS data. \textit{Right}: The redshift distribution of WFSS selected HAEs. The two prominent peaks at $z\approx4.41$ and $z\approx 5.19$ indicate the presence of protoclusters. }
    \label{fig:distribution}
\end{figure*}

We construct the final HAE catalog from the preliminary emission line catalog (\S\ref{sec:emission_line_selection}). We identify clumpy HAEs with clump separations $< 0.5\arcsec$ and velocity offsets $< 500$ \si{km.s^{-1}}. These clumpy galaxies may be detected as multiple components in the JADES catalog, with the spectra of the clumps overlapping along the dispersion direction. We have identified 62 clumpy HAEs at $z\approx3.75-6$. For each system, we merge the segmentations of the clumps in the JADES GOODS-N segmentation map into a single segment, re-measure their photometry using \textsc{Photutils} \citep{photutils} based on the merged segmentation, and re-extract the spectra for these sources using the updated morphological parameters.  The final HAE catalog contains \NHAEtot\ sources at $3.75<z<6.6$ with the \ha\ line detected at S/N $>5$.  We then exclude AGNs based on broad-line features identified in grism or NIRSpec data \citep[e.g.,][]{Maiolino2023, Matthee2024_lrd, Rinaldi2024, Zhang2025}, or objects exhibiting V-shaped continua characteristic of ``little red dots" \citep[e.g.,][]{Matthee2024_lrd, Greene2024}.  There are a total of 28 AGNs and 908 star-forming galaxies at $3.75<z<6.6$. The AGN catalog and further studies on their properties are beyond the scope of this paper and will be detailed in \cite{Zhang2025} and \cite{Lin_AGN_clustering}.  This work focuses on \NHAE\ HAEs at $3.75 < z < 6$, and the following analysis is based on two bins within this range ($3.75<z<5$, $5<z<6$).  Only 20 HAEs are identified at $z = 6 - 6.6$. Although we define $z = 6 - 6.6$ as a separate bin due to the rapid evolution of galaxies at $z > 6$ \citep[e.g.,][]{GSun2023, Endsley2024, Endsley2024b}, this redshift range is not included in our study due to the limited sample size. Figure \ref{fig:distribution} shows the spatial and redshift distribution of HAEs at $3.75 < z < 6$.

\begin{figure*}
    \centering
    \includegraphics[width=0.49\linewidth]{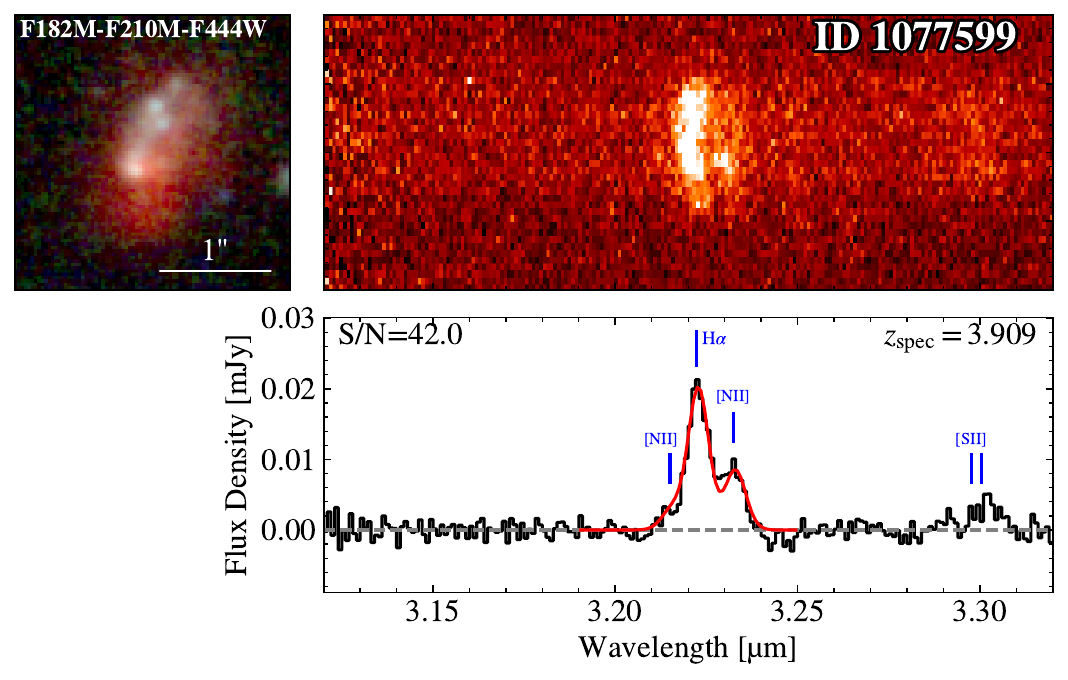}
    \includegraphics[width=0.49\linewidth]{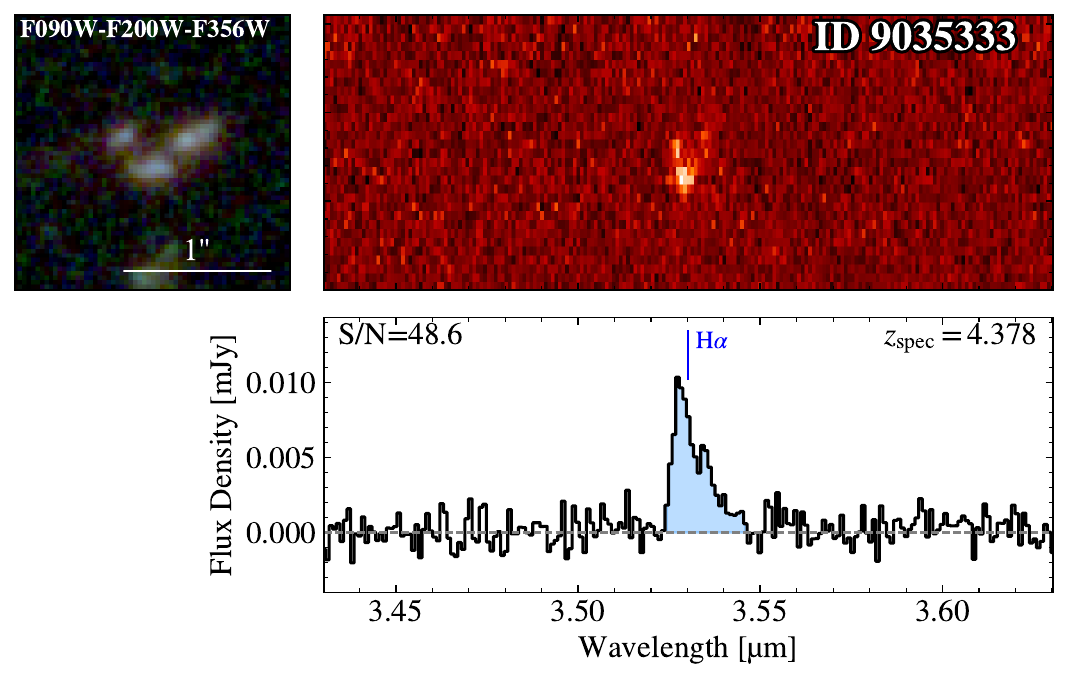}
    \caption{Two examples of HAEs in our sample. The top right panel shows the 2D spectrum and the bottom panel presents the optimally extracted spectrum.  \textit{Left}: A bright HAE at $z=3.909$ with prominent [\ion{N}{2}] and [\ion{S}{2}]. The best-fit line profile, which convolves the intrinsic \ha\ and [\ion{N}{2}] profiles with the LSF, is shown as the red line. \textit{Right}: A clumpy galaxy consisting of three components. This galaxy was identified as three separate objects in the JADES DR3 catalog \citep{JADES_DR3}. We have combined them into a single segmentation, updated the photometry and morphology parameters, and re-extracted the spectrum. The line profile is not well described by a single Gaussian, so we integrate the line flux over the area highlighted in the blue-shaded region.}
    \label{fig:HAE_example}
\end{figure*}

We use the originally extracted 1D spectra, rather than the median-filtered spectra, to measure the line flux. We mask regions $\pm$ 300\,\AA\ around the H$\alpha$ emission line and fit a spline function to the extracted 1D spectrum within this range to remove the local continuum. For sources that are severely contaminated by nearby bright sources, we use 1D spectra extracted from the median-filtered 2D spectrum. We measure the \ha\ line flux by fitting the 1D spectra with Gaussian profiles convolved with the line-spread functions \citep[LSF,][]{Sun_LSF}. For unresolved emission lines, we rescale the LSFs to match the lines. For galaxies with prominent [\ion{N}{2}] emission lines, we simultaneously fit the \ha\ + [\ion{N}{2}] profiles, convolved with the LSFs. Clumpy galaxies may have morphologically broadened profiles that cannot be well described by a Gaussian. In these cases, we integrate the emission line region in 1D spectra to obtain the total flux. We show two examples of our HAEs in Figure \ref{fig:HAE_example}.

We compared our \ha\ emitter catalog with the $z > 4.9$ HAE catalog from \cite{CoveloPaz2024}, which is based on F444W Grism R data from the FRESCO program. \xj{Different from the selection method in \citet{CoveloPaz2024}, our redshift algorithm utilizes both photometric redshifts and multiple expected emission lines. Specifically, for HAEs at $z > 5$, when [\ion{O}{3}] falls within the F356W spectral coverage, the algorithm combines the photometric redshift with detections of [\ion{O}{3}] and \ha\ in the F356W grism to determine the redshift and its confidence.} The catalog from \cite{CoveloPaz2024} contains 318 HAEs identified in the GOODS-N field, of which 258, i.e., about 81\%, are included in our catalog. Among the remaining sources, eight are classified as AGNs based on broad-line features identified in grism or NIRSpec spectroscopy \citep{Maiolino2023, Zhang2025, Rinaldi2024} or their V-shaped SEDs \citep{Lin_AGN_clustering}. Additionally, two sources are part of our [\ion{O}{3}] emitter catalog but have low S/N in \ha\, and thus are not included in this study. Fifteen of the remaining sources have most probable redshifts of $z<3.5$. For the rest, either their spectral lines are assigned to other sources along the dispersion direction that have better phot-$z$ matching the line positions, or their estimated spec-$z$ suggests that the expected [\ion{O}{3}] lines should fall within the F356W grism coverage, but are not detected. In our catalog, there are 367 $z > 4.9$ HAEs with expected \ha\ lines in the F444W grism. Among these, 242 have been confirmed in \cite{CoveloPaz2024}. Among the remaining sources, 13 exhibit significant lines in the F356W grism but have \ha\ located at the edge of F444W, placing them outside the scope of the F444W-selected HAEs in their catalog. The rest either have reliable photometric redshifts with $|z_{\rm phot} - z_{\rm spec}| < 0.05$ or show [\ion{O}{3}] in the F356W grism.   \cite{CoveloPaz2024} has provided a solid foundation for identifying HAEs in the GOODS-N field, and our catalog further complements their findings by identifying additional sources and refining the redshift estimates.

\begin{figure*}[htbp!]
    \centering
\includegraphics[width=0.75\textwidth]{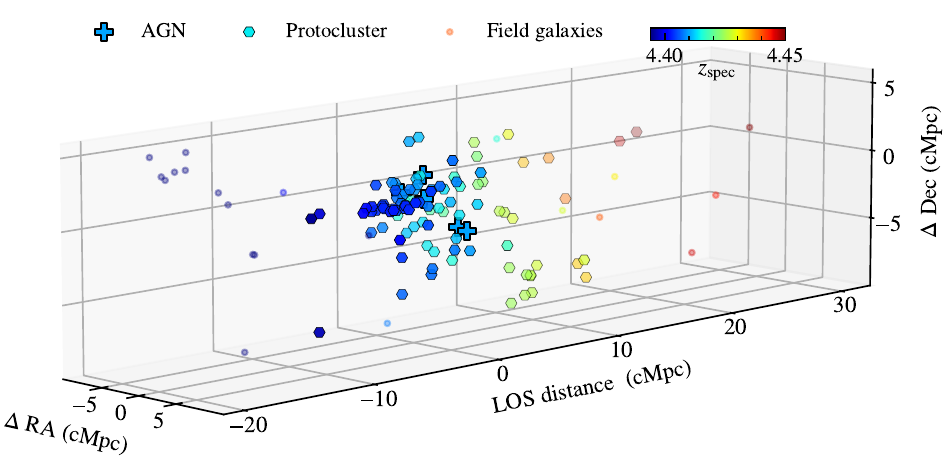}
\includegraphics[width=0.75\textwidth]{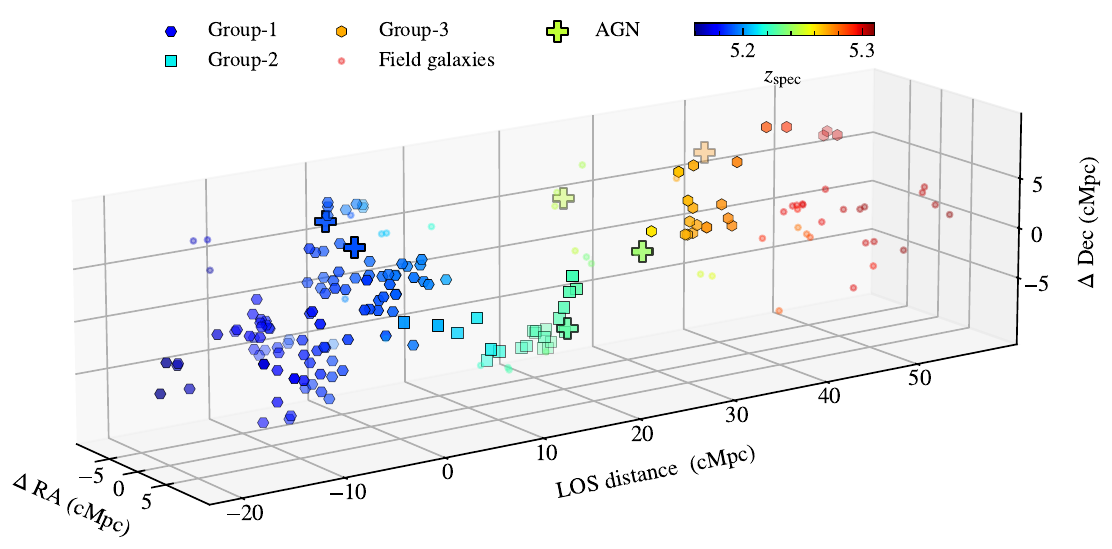}
    \caption{The 3D large-scale structure of the protocluster at $z\approx4.41$ (top panel) and $z\approx5.19$ (bottom panel). The coordinates of the $z\approx4.41$ structure are with respect to (RA, DEC, $z$)=(189.21355, 62.24861, 4.41), and the coordinates of the $z\approx5.19$ structure are
    with respect to (RA, DEC, $z$)=(189.20403, 62.23787, 5.195). All sources are color-coded by their redshifts, with the plus indicating AGNs, hexagons/squares indicating the protocluster members, and dots indicating field galaxies. }
    \label{fig:protocluster_structure}
\end{figure*}

\subsection{Protoclusters at $z=4.41$ and $z=5.19$}

We detect two prominent redshift peaks in the HAE redshift distribution (right panel of Figure \ref{fig:distribution}) at $z\approx4.41$ and $z\approx5.19$, suggesting significant overdensities at these redshifts.  The two redshift peaks have also been confirmed in \cite{CoveloPaz2024}. We run a Friends-of-Friends algorithm with projected separations smaller than 500 physical kpc and line-of-sight (LOS) velocity offsets below 500 km s$^{-1}$ following \citet{Helton2024}.  We identify a protocluster at $z=4.41$ containing 98 galaxies and another at $z=5.19$ associated with 144 galaxies. The $z=4.41$ protocluster exhibits a filamentary structure spanning 10 cMpc on the projected sky plane and 25 cMpc along the LOS (Figure \ref{fig:protocluster_structure}).  The $z\approx5.19$ protocluster, associated with the brightest submillimeter galaxy in the Hubble Deep Field, HDF850.1, was first reported by \citet{Walter2012} and recently studied by JWST through \cite{Sun2024}, \citet{Helton2024} and \cite{Herard_Demanche_2023}.  This protocluster contains three groups of galaxies, with 100 galaxies at  $z\approx5.19$,  23 galaxies at $z\approx5.227$, and 21 galaxies at $z\approx5.269$. Its structure fills the WFSS footprint on the projected sky plane ($\sim$15 cMpc) and extends $\sim$70 cMpc along the LOS (Figure  \ref{fig:protocluster_structure}). 
The $z\approx4.41$ protocluster contains five low-luminosity AGNs, identified through broad \ha\ line (\citealt{Zhang2025}, using the NIRCam grism, and \citealt{Maiolino2023} with NIRSpec) and V-shaped SEDs \citep{Lin_AGN_clustering}. The $z\approx5.19$ protocluster hosts six AGNs. A detailed analysis of the environments surrounding these AGNs is presented in a companion paper \citep{Lin_AGN_clustering}.

\subsection{The observed \ha\ Luminosity function}\label{sec:HaLF_measure}

\begin{figure*}[tbp!]
    \includegraphics[width=\linewidth]{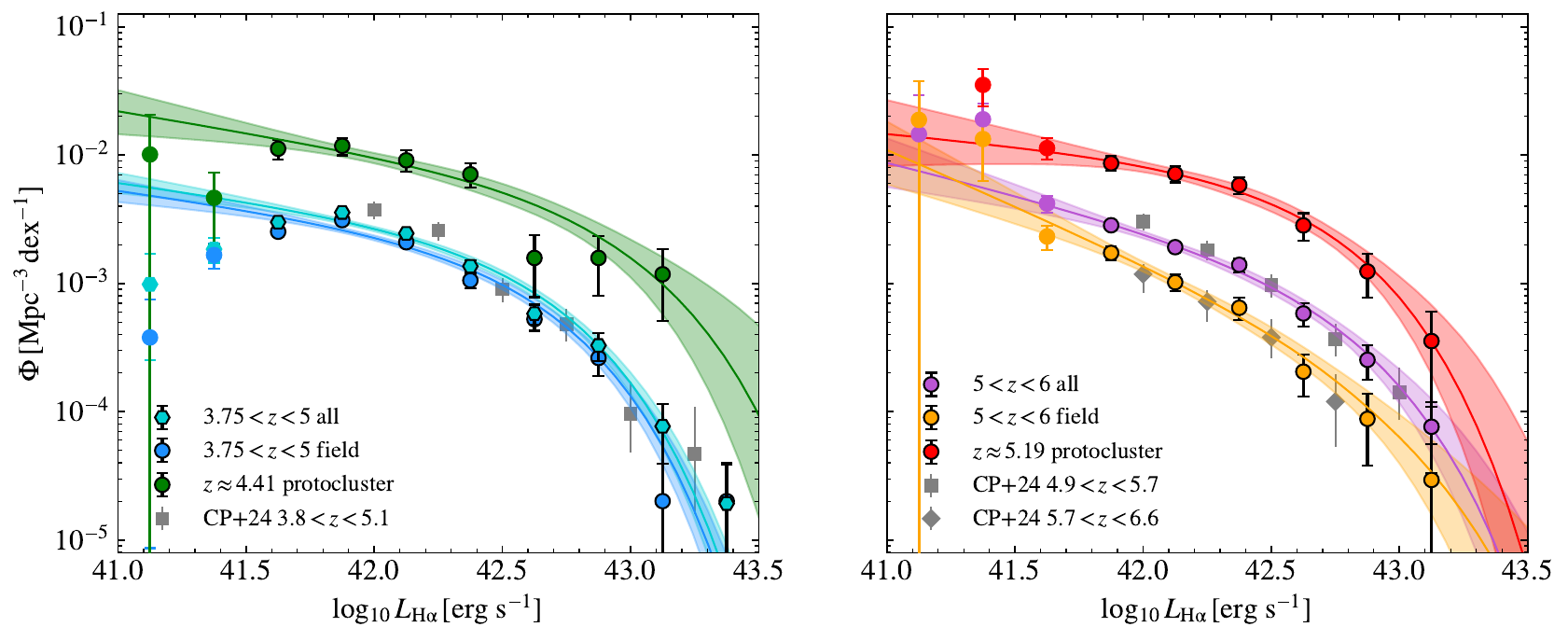}
    \caption{The observed \ha\ LFs using the direct 1/$V_{\rm max}$ method. \textit{Left}: The observed \ha\ LFs at $3.75<z<5$. The cyan dots represent the overall \ha\ LF at $3.75<z<5$ including all HAEs within this redshift range. The green dots represent the \ha\ LF within $z = 4.39 - 4.45$, dominated by the $z \approx 4.41$ protocluster. The blue dots are the LF of field galaxies by excluding the protocluster regions. \xj{The uncertainties of the LF measurements are estimated using bootstrapping.} The data points without black edges have average completeness values of $< 0.7$ and are not included in the Schechter function fit. The best-fit Schechter function and its uncertainty are represented by solid lines and color-shaded regions.  We also show the overall \ha\ LFs at $3.8<z<5.1$ in \cite{CoveloPaz2024} as the gray squares for reference. \textit{Right:} \ha\ LFs at $5<z<6$. The overall \ha\ LF is shown as the purple dots, the one dominated by the $z=5.19$ protocluster is shown in red and the one including only field galaxies is shown in orange.  The gray squares and diamonds are the overall \ha\ LFs at $z=4.9-5.7$ and $z=5.7-6.6$ in \cite{CoveloPaz2024}, respectively.
     \label{fig:HaLF}
     }
\end{figure*}

We calculate the luminosity functions (LFs) by first computing the total survey volume of HAEs at $z=3.75-6$. We construct a series of root mean square (RMS) mosaicked maps for emission lines in the 3.1--5.0\,\micron\ range, with intervals of 0.005\,\micron. For the overlapping wavelength range of the F356W and F444W grism, we set 3.925\,\micron\ as the boundary: at $\lambda > 3.925\,\mu\text{m}$ the F444W grism has higher sensitivity where emission lines can be more easily identified. The RMS maps are computed based on the spectral tracing and grism dispersion
models using the continuum-removed WFSS \texttt{rate} files.  The maximum sky area for an emission line is defined as the region on the RMS map where the RMS value is below the maximum RMS for its identification. To determine the maximum RMS for each emitter, we create 2D spectra by inserting mock emission lines with specified line fluxes and RMS noise levels. We then apply the emission line detection algorithm described in \S\ref{sec:emission_line_selection}. By gradually increasing the RMS noise, we identify the specific noise level at which the line can no longer be detected across 1,000 trials and set this as the maximum RMS for line detection. For HAEs over a given redshift range, the maximum survey volume ($V_{\rm max}$) is calculated by integrating the maximum sky areas across the \ha\ wavelength range. 

We \xj{primarily} adopt two approaches to estimate the \ha\ LFs: the direct $1/V_{\rm max}$ method \citep{Schmidt1968} and the maximum likelihood estimator (MLE) method \citep{Sandage1979}. \xj{We further employ the non-parametric Lynden-Bell's $C^-$ method \citep{Lynden-Bell1971}. The Lynden-Bell's $C^-$ method is an unbinned maximum likelihood estimator of the cumulative luminosity function for flux-limited truncated data. We present LF measurements using the $C^-$ method in Appendix~\ref{sec:appendix_lf_cm}. The results are fully consistent with those using the $1/V_{\rm max}$ and the MLE method.} The $1/V_{\rm max}$ and MLE methods are described below.

For direct 1/$V_{\rm max}$ method, the LF is computed as:
\begin{equation}\label{eq:one_over_Vmax}
\Phi(L)=\frac{1}{d \log L} \sum_i \frac{1}{C_i V_{\max , i}},
\end{equation}
where $L$ is the \ha\ luminosity of each LF bin, $C_i$ accounts for the completeness correction, and $V_{\max,i}$ is the maximum survey volume for the $i-$th HAEs. \xj{We compute $V_{\max,i}$ following the method described by \cite{Sun2023}. This approach integrates the maximum observable survey area where the emission line is detectable, given the RMS noise in the 2D spectral images and the detection threshold.} To measure $C_i$, we generate a completeness model as a function of line wavelength, line flux, and RMS values. In brief, we generate mock spectra with specific line wavelengths, fluxes, and RMS values. Then we conduct 1,000 line detection trials. The fraction of successful detections across the trials, or the recovery rate of emission lines, defines the completeness of emission lines at fixed line flux and RMS values. Appendix \ref{sec:completeness_function} presents more details about the completeness function. For each HAE, we obtain an array of completeness by interpolating its line flux and the RMS values within its maximum survey volume into the completeness grids. We then determine $C_i$ by averaging the completeness array. \xj{The uncertainties of the LFs are estimated through 1000 bootstrap resamplings.}

We calculate the \ha\ LFs in two redshift bins: $z=3.75-5$ and $z=5-6$. We first measure the overall \ha\ LFs by including all HAEs within these bins.  We then calculate the \ha\ LFs for galaxies in the protoclusters and fields, respectively. For the $z=4.41$ protocluster, we restrict the redshift range to $z=4.39-4.45$, where over 91\% of galaxies are protocluster members. To estimate the \ha\ LF in the field, we compute the LF by masking the emitters and survey volume within this range. For the $z=5.19$ protocluster, we limit the redshift range to $z=5.15-5.29$, where 78\% of galaxies are protocluster members. \xj{Note that the redshift ranges of the two protoclusters are defined by the minimum and maximum redshifts of their identified members. The different fractions of protocluster member galaxies arise from the galaxy distribution within these intervals and the relaxation of boundaries between protocluster members and surrounding galaxies, which in turn affect the outcomes of the Friends-of-Friends algorithm.} However, since protoclusters dominate these redshift ranges, this difference does not affect the subsequent analysis. The overall \ha\ LFs, as well as the LFs within the protoclusters and in the fields, are shown in Figure \ref{fig:HaLF}.  The measurements are listed in Table \ref{tab:HaLF}.

We model the \ha\ LFs in Equation \ref{eq:one_over_Vmax} using the Schechter function \citep{Schechter1976}:
\begin{equation}\label{eq:Schechter}
     \Phi(L) dL =  \Phi_* \left(\frac{L}{L_*}\right)^\alpha e^{-L/L_*}  \frac{dL}{L_*}.
\end{equation}
We run Markov chain Monte Carlo (MCMC) through \textsc{emcee} \citep{emcee} with a flat prior and obtain the posterior probability distribution of  $\Phi_*$, $L_*$ and $\alpha$.  

For the MLE method, we parameterize the LFs using the Schechter function in Equation \ref{eq:Schechter}. For a given parametric LF description $\Phi$, the probability of detecting a galaxy with luminosity $L_i$ is given by:  
\begin{equation}\label{eq:MLE_P}
P\left(L_i\right) = \frac{\Phi\left(L_i\right) \cdot V_{\text{eff},i}}{\int_{L_{\text{lim}}}^{\infty} \Phi(L) \cdot V_{\text{eff}}(L) \, \mathrm{d}L},
\end{equation}  
where $\Phi(L_i)$ is the expected LF value at H$\alpha$ luminosity $L_i$, given the Schechter parameters $\Phi_*$, $L_*$, and $\alpha$ as defined in Equation \ref{eq:Schechter}. $V_{\text{eff}}$ represents the effective survey volume. For individual galaxies, $V_{\text{eff},i} = C_i V_{{\rm max},i}$. To obtain $V_{\text{eff}}(L)$, we interpolate $V_{\text{eff},i}$ as a function of H$\alpha$ luminosity using a spline function (Appendix \ref{sec:Appendix_MLE_result}). We set the integration limit $L_{\rm lim}$ to the minimum H$\alpha$ luminosity of the input galaxies.

The log-likelihood function for the full sample can then be computed as:
\begin{equation}\label{eq:MLE_lh}
\ln \mathcal{L}=\sum_{i=1}^N \ln P\left(L_i\right)
\end{equation}
We use MCMC to maximize the log-likelihood function and find the best-fit Schechter parameters. In Equation~\ref{eq:MLE_P}, $\Phi_*$ is canceled out by the normalization, so it cannot be well constrained by maximizing Equation~\ref{eq:MLE_lh} alone. We re-compute $\Phi_*$ for each MCMC chain using the following expression:
\begin{equation}
\Phi_{\star} = \frac{N}{\int_{L_{\rm lim}}^{\infty} \Phi(L) / {\Phi_*^\prime} \cdot V_{\text{eff}}(L) \, \mathrm{d}L},
\end{equation}
where $\Phi_*^\prime$ is the original value of $\Phi_*$ output by each MCMC chain, and $N$ is the total number of galaxies. Similar to the direct $1/V_{\rm max}$ method, we use the MLE method to estimate the H$\alpha$ LFs for the full sample at $3.75 < z < 5$ and $5 < z < 6$, the LFs dominated by the protoclusters at $z \approx 4.41$ and $z = 5.19$, and the field LFs by masking the volume occupied by the protoclusters.

\begin{table*}
	\begin{center}
		\begin{tabular}{cccc|ccc}
			\hline
            \hline
			   & $\log (\phi_*/{\rm Mpc^{-3}\, dex^{-1}})$ & $\log (L_*/{\rm erg\,s^{-1}})$ & $\alpha$ &  $\log (\phi_*/{\rm Mpc^{-3}\, dex^{-1}})$ &  $\log (L_*/{\rm erg\,s^{-1}})$ &  $\alpha$ \\
            \hline
			 & \multicolumn{3}{c|}{All $3.75<z<5.00$} &  \multicolumn{3}{c}{All $5<z<6$}   \\
             \hline 
			Direct 1/$V_{\rm max}$ & $-3.00^{+0.09}_{-0.10}$ & $42.62^{+0.07}_{-0.06}$ & $-1.26^{+0.09}_{-0.09}$ & $-3.27^{+0.22}_{-0.36}$ & $42.75^{+0.26}_{-0.15}$ & $-1.49^{+0.20}_{-0.21}$ \\
			MLE & $-2.92^{+0.10}_{-0.12}$ & $42.55^{+0.09}_{-0.08}$ & $-1.18^{+0.13}_{-0.12}$ & $-3.35^{+0.30}_{-0.56}$ & $42.82^{+0.42}_{-0.22}$ & $-1.49^{+0.27}_{-0.26}$ \\
            \hline
			 & \multicolumn{3}{c|}{Field $3.75<z<5.00$}   & \multicolumn{3}{c}{Field $5<z<6$}  \\
             \hline
			Direct 1/$V_{\rm max}$ & $-3.07^{+0.10}_{-0.12}$ & $42.61^{+0.09}_{-0.07}$ & $-1.27^{+0.10}_{-0.11}$ & $-4.02^{+0.51}_{-0.84}$ & $42.95^{+0.57}_{-0.31}$ & $-1.87^{+0.30}_{-0.23}$ \\
			MLE & $-2.92^{+0.10}_{-0.12}$ & $42.48^{+0.09}_{-0.08}$ & $-1.11^{+0.14}_{-0.14}$ & $-4.03^{+0.63}_{-0.87}$ & $42.98^{+0.62}_{-0.41}$ & $-1.81^{+0.40}_{-0.26}$ \\
            \hline
			 & \multicolumn{3}{c|}{Protocluster $4.39<z<4.43$} &  \multicolumn{3}{c}{Protocluster $5.15<z<5.29$}    \\
             \hline 
			Direct 1/$V_{\rm max}$ & $-2.65^{+0.28}_{-0.43}$ & $42.95^{+0.45}_{-0.24}$ & $-1.33^{+0.22}_{-0.19}$ & $-2.47^{+0.20}_{-0.36}$ & $42.66^{+0.29}_{-0.16}$ & $-1.17^{+0.27}_{-0.29}$ \\
			MLE & $-2.49^{+0.23}_{-0.40}$ & $42.78^{+0.40}_{-0.21}$ & $-1.19^{+0.22}_{-0.22}$ & $-2.70^{+0.33}_{-0.60}$ & $42.86^{+0.54}_{-0.27}$ & $-1.36^{+0.32}_{-0.26}$ \\
			\hline
		\end{tabular}
	\end{center}
    \caption{The best-fit Schechter parameters for the overall \ha\ LFs at the two redshift bins, and LFs within the protoclusters and in the fields. The Schechter function is fitted using both the direct 1/$V_{\rm max}$ method and the MLE method.} 
    \label{tab:Schechter}
\end{table*}

\begin{figure}[h!]
    \centering
    \includegraphics[width=\linewidth]{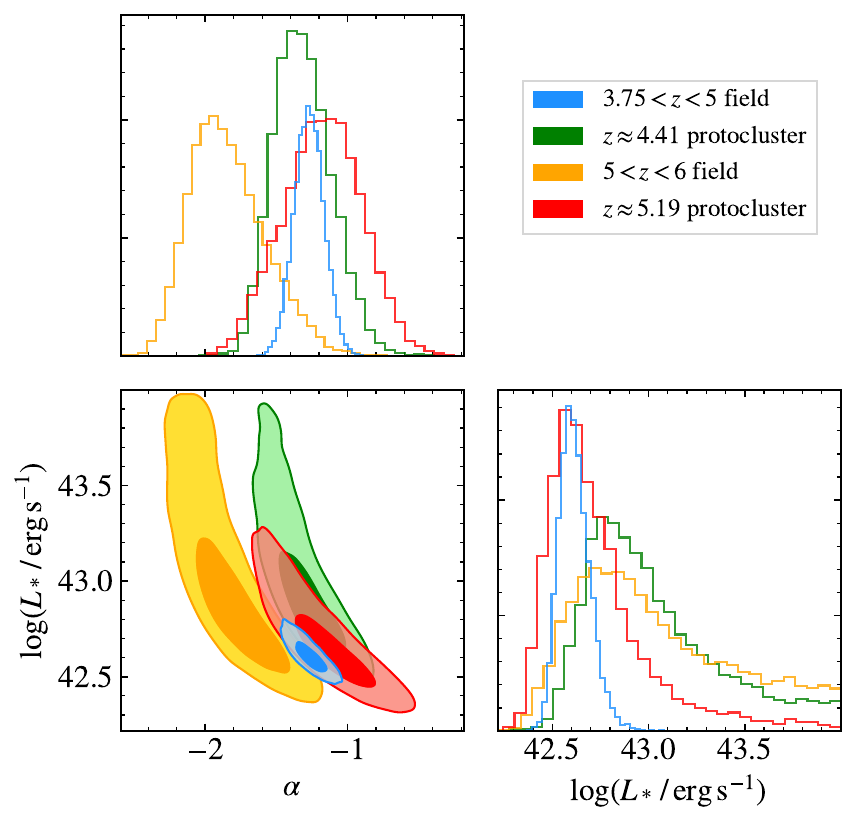}
    \caption{The posterior probability distribution of the faint-end slope ($\alpha$) and characteristic luminosity ($L_*$) for the field and protocluster \ha\ LFs in the redshift ranges $3.75 < z < 5$ and $5 < z < 6$. The posterior probability is obtained by fitting the H$\alpha$ LFs from the direct $1/V_{\rm max}$ method using MCMC. The deep and light color-shaded regions represent the 1$\sigma$ and 2$\sigma$ ranges of the 2D probability distribution, respectively.}
    \label{fig:HaLF_param_prob}
\end{figure}

We list the best-fit Schechter parameters from both methods in Table \ref{tab:Schechter}. The two methods yield consistent results within 1$\sigma$. In the following, we will primarily focus on the LF constraints from the direct $1/V_{\rm max}$ method, while detailed results from the MLE method, which are in good agreement with those of the direct $1/V_{\rm max}$, are presented in the Appendix \ref{sec:Appendix_MLE_result}.

 At $3.75 < z < 5$, the \ha\ LFs of galaxies within protoclusters and in the field show consistent slopes of $\alpha\approx -1.3$. The characteristic luminosity  ($L_*$) of protocluster galaxies is 0.34 dex higher than that of field galaxies. At $5 < z < 6$, the protocluster \ha\ LF has a faint-end slope of $\alpha \approx -1.17^{+0.27}_{-0.29}$. In contrast, the field LF has a faint-end slope of $\alpha \approx -1.87^{+0.30}_{-0.29} $, which is steeper than the protocluster LF by 0.7. The faint-end observed \ha\ LF in the $z = 5.19$ protoclusters (red points in the right panel of Figure \ref{fig:HaLF}) clearly exhibits a suppressed shape in luminosity bins with $L_{\rm H\alpha} < L_*$ compared to the field LF at $z = 5 - 6$ (orange points in the right panel of Figure \ref{fig:HaLF}). We show the posterior probability distribution of $\alpha$ and $L_*$ in Figure \ref{fig:HaLF_param_prob}. The posterior probability distribution of $\alpha$ for the $z = 5.19$ protocluster LF occupies the same region as that of the protocluster LF at $z = 4.41$ and the field LF at $3.75 < z < 5$ within 1$\sigma$ uncertainties. In contrast, the field LF at $z = 5 - 6$ deviates from them by nearly 2$\sigma$. We conclude that the field and protocluster galaxies at $z = 3.75 - 5$, as well as the protocluster galaxies at $z = 5.19$, have similar faint-end slopes that are consistent with each other within 1$\sigma$, whereas the field galaxy LF at $z = 5 - 6$ exhibits a steeper slope.  We discuss the implications of the LF shapes on the star formation in protoclusters and field galaxies in \S\ref{sec:implication_sf}.

\subsection{The observed cosmic star formation rate density}
We integrate the H$\alpha$ LF down to $L_{\rm H\alpha, \min} = 10^{41}$ erg s$^{-1}$, a limit similar to that in \cite{CoveloPaz2024}, to derive the observed H$\alpha$ luminosity density:
\begin{equation}
    \rho_{\rm H\alpha} = \int_{L_{\rm H\alpha, \min}} \Phi(L) L dL
\end{equation}
We calculate the observed cosmic star formation rate density ($\rho_{\rm SFR}$) by converting $\rho_{\rm H\alpha}$ assuming the \cite{Chabrier2003} initial mass function (IMF):
\begin{equation}\label{eq:Ha_SFR}
    \log (\rho_{\rm SFR}/M_\odot\,{\rm yr}^{-1}) = \log (\rho_{\rm H\alpha}/{\rm erg\,s^{-1}}) - 41.35.
\end{equation}
Note that the $\rho_{\rm SFR}$ here is directly derived from observed \ha\ luminosity densities and has not been corrected for dust attenuation. We do not apply a dust correction in this study, because the dust attenuation laws across galaxies with different stellar populations remain uncertain in the high-redshift Universe \citep[e.g.,][]{Sanders2024, Markov2024}.

We estimate the contribution of $\rho_{\rm SFR}$ from the protoclusters at $3.75<z<5$ and $5<z<6$. We integrate the protocluster LFs ($\Phi_{\rm protocluster}$), multiply the results by the volumes that the protoclusters occupy ($V_{\rm protocluster}$), and then divide them by the total volumes across the redshift range ($V_{\rm tot}$). \xj{This defines the $ \rho_{\rm SFR} $ contributed by protoclusters over a specified redshift interval. In our case, it is the effective $ \rho_{\rm SFR} $ contributed by the $ z \approx 4.41 $ protocluster to $ 3.75 < z < 5 $ and by the $ z \approx 5.19 $ protocluster to $ 5 < z < 6 $. We obtain }
\begin{equation}
    \rho_{\rm H\alpha, protocluter} = \frac{V_{\rm protocluster}}{V_{\rm tot}}\int_{L_{\rm H\alpha, \min}} \Phi_{\rm protocluster}(L) L dL.
\end{equation}
Then we estimate the star formation rate density contributed by protoclusters ($\rho_{\rm SFR, protocluster}$) using Equation \ref{eq:Ha_SFR}.  On the other hand, $\rho_{\rm H\alpha}$ contributed by field galaxies are described as
\begin{equation}
    \rho_{\rm H\alpha, field} = \frac{V_{\rm field}}{V_{\rm tot}}\int_{L_{\rm H\alpha, \min}} \Phi_{\rm field}(L) L dL.
\end{equation}
where $V_{\rm field} = V_{\rm tot} - V_{\rm protocluster}$.

\begin{figure}[htbp!]
    \centering
    \includegraphics[width=1\linewidth]{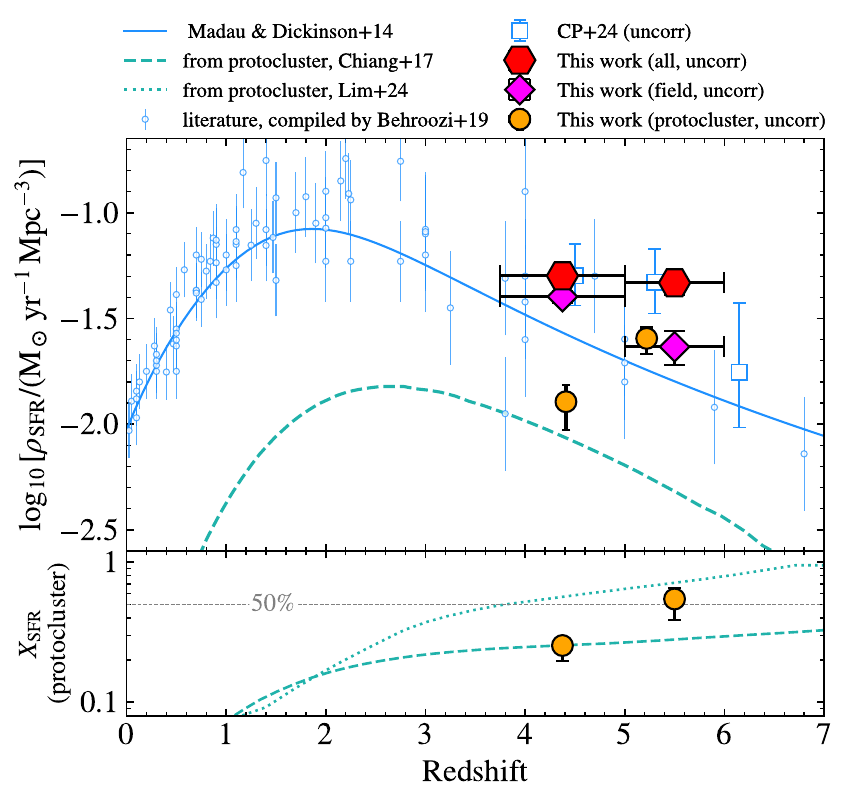}
    \caption{\textit{Top panel}: The observed evolution of the cosmic SFR density. We show the literature measurements using multiple tracers \citep[UV, IR, radio, etc., compiled by ][see references therein]{Behroozi2019}. All the measurements and the \cite{Madau2014} parametrization are corrected to the \cite{Chabrier2003} IMF. The light blue squares are derived by integrating the Schechter function of the observed \ha\ LFs in \cite{CoveloPaz2024} down to $L_{\rm H\alpha, \min}=10^{41}$ erg s$^{-1}$, without correcting for dust attenuation. We illustrate the $\rho_{\rm SFR}$ contributed by protoclusters predicted in \cite{Chiang2017} as the green dashed lines. \textit{Bottom panel}: We show the predicted fraction of $\rho_{\rm SFR}$ contributed by protoclusters, $X_{\rm SFR}$, from \cite{Chiang2017} and \cite{Lim2024} as green dashed and dotted lines, respectively.
    }
    \label{fig:csfh}
\end{figure}

The  cosmic star formation rate densities $\rho_{\rm SFR}$ over $3.75<z<5$ and $5<z<6$ are displayed in Figure \ref{fig:csfh}.  We also show the fraction of star formation rate density contributed by the protoclusters. We find that, at $3.75<z<5$, $\rho_{\rm SFR}$ is  0.27 dex higher than the parametrization of \cite{Madau2014} and is 0.47 dex higher at $5<z<6$. The high $\rho_{\rm SFR}$ values are consistent with \cite{CoveloPaz2024}. The weak evolution from $z\approx 4.5$ to $z\approx5.5$, unlike the \cite{Madau2014} curve, is primarily due to the fact that our observed $ \rho_{\rm SFR} $ has not been corrected for dust attenuation. Dust attenuation in star-forming galaxies becomes more significant at $ z \approx 4 $ than at $ z \approx 5 $, meaning that the correction factor would be higher at $ z \approx 4 $ than at $ z \approx 5 $.  \xj{As our HAE sample and measured \ha\ LFs are in close agreement with those of \citet{CoveloPaz2024} (\S\ref{sec:hae_sample} and Figure \ref{fig:HaLF}), we adopt their LFs as a reference for assessing the impact of dust attenuation. When the observed and dust-corrected \ha\ LFs in \citet{CoveloPaz2024} are integrated down to $L_{\rm H\alpha,min} = 10^{41}\,{\rm erg\,s^{-1}}$ as in this work, the observed $\rho_{\rm SFR}$ corresponds to approximately 37\% of the dust-corrected $\rho_{\rm SFR}$ at $z \approx 4.45$ and approximately 43\% at $z \approx 5.30$, respectively.}

When considering the field galaxies alone, we find that the $\rho_{\rm SFR, field}$ follows the redshift evolution of the \cite{Madau2014} curve, but it remains above the curve by 0.19 and 0.23 dex at $3.75 < z < 5$ and $5 < z < 6$, respectively.

\begin{table}[h!]
\centering
	\begin{tabular}{ccc}
    \hline
    & \multicolumn{2}{c}{$ \log(\rho_{\rm SFR} /M_\odot\,{\rm yr}^{-1}\,{\rm Mpc}^{-3})$} \\
    \hline
    Redshift & $3.75<z<5$ & $5<z<6$ \\
    All & $-1.30^{+0.03}_{-0.02}$ & $-1.33^{+0.06}_{-0.04}$ \\
    Field & $-1.40^{+0.03}_{-0.03}$ & $-1.63^{+0.08}_{-0.07}$ \\
    Protocluster & $-1.89^{+0.14}_{-0.08}$ & $-1.59^{+0.07}_{-0.05}$ \\
    $X_{\rm SFR}$ (protocluster) & $0.25^{+0.06}_{-0.03}$ & $0.55^{+0.16}_{-0.11}$ \\
    \hline
	\end{tabular}
    \caption{The observed cosmic SFR density $\rho_{\rm SFR}$ contributed by all the HAEs, and HAEs in the protoclusters and the fields (uncorrected for dust).  $\rho_{\rm SFR}$ is integrated down to $L_{\rm H\alpha} = 10^{41}$ \si{erg\,s^{-1}}. $X_{\rm SFR}$(protocluster) is the fraction of $\rho_{\rm SFR}$ contributed by the protoclusters within the redshift bin. }
\end{table}

Our results suggest that the unobscured SFR density is significantly higher than empirical predictions. On the other hand, a substantial fraction of star formation could occur in dust-enshrouded environments \cite[e.g.,][]{Casey2018, Fujimoto2024}. Recently, \cite{Sun2025} measured the dust-obscured SFR density at $z = 4 - 6$ across 25 independent sightlines \citep[``A SPectroscopic survey of biased halos In the Reionization Era", ASPIRE;][]{Wang2023}, covering a total area of $\sim 35$\,arcmin$^2$. Their sample consists of eight star-forming galaxies detected in the ALMA 1.2-mm continuum map, with \ha\ or [\ion{O}{3}] emission lines confirmed through NIRCam grism observations. They report an obscured SFR density of $\log(\rho_{\rm SFR, IR}/M_\odot {\rm yr^{-1}} {\rm Mpc^{-3}}) = -1.52^{+0.14}_{-0.13}$ at $z = 4 - 6$, which corresponds to $59^{+26}_{-18}$\% of the H$\alpha$-derived observed $\rho_{\rm SFR}$ at $3.75 < z < 5$ and $67^{+36}_{-23}\%$ at $5 < z < 6$. This suggests that the majority ($\sim 62\%$) of total star formation at $z=4-6$ is obscured at the rest-frame UV wavelength, while the unobscured component still contributes a significant fraction ($\sim 38\%$) of the total.  \xj{This conclusion is consistent with the dust-corrected $\rho_{\rm SFR}$ obtained from \ha\ LFs. The comparison between observed and dust-corrected \ha\ LFs in \citet{CoveloPaz2024} indicates that, when integrated to the same limit as in this work, approximately 63\% and 57\% of the star formation at $z \approx 4.45$ and $z \approx 5.30$, respectively, is obscured by dust.}

The protocluster contributes about $25^{+6}_{-3}\%$ of the total $\rho_{\rm SFR}$ over $3.75<z<5$, based on the ratio of $\rho_{\rm SFR, protocluster}$ to the total $\rho_{\rm SFR}$. The actual contribution could be higher, as we do not account for other smaller overdensities/galaxy associations \citep{Helton2024}. At $5<z<6$, the contribution of protoclusters to $\rho_{\rm SFR}$ rises to $55^{+16}_{-11}\%$ (see also \citealt{Sun2024}). It far exceeds the prediction of $\sim 28\%$ from simulations by \cite{Chiang2017} at $z \sim 5.5$. On the other hand, \cite{Lim2024} predicts about 62\% star formation occurs in protoclusters at $z\sim 4.5$ and 74\% at $z\sim 5.5$. Both predicted fractions are higher than our measured values. Though qualitatively consistent, the discrepancy may be attributed to cosmic variance and the limitations in the statistics of protoclusters at $z>4$. Nevertheless, our $\rho_{\rm SFR}$ estimates suggest that protoclusters or overdensities could contribute significantly to the overall cosmic star formation history at $z > 4$, given the ubiquity of overdensities observed in JWST surveys. In other words, a large fraction of star formation at $z > 4$ occurs in protoclusters.

\subsection{Implications for star formation in protoclusters and field galaxies at $z>5$}\label{sec:implication_sf}

As discussed in \S\ref{sec:HaLF_measure},   the faint-end slopes $z=3.75-5$ of the \ha\ LFs in protoclusters and the field are similar. At $z = 5-6$, the faint-end slope of the protocluster \ha\ LF shows little evolution compared to those observed at lower redshifts. In contrast, the field \ha\ LF evolves more rapidly from $z = 5-6$ to $z = 3.75-5$, with the slope of the $z = 5-6$ field \ha\ LF being steeper by 0.7. This result is broadly consistent with the trend of [\ion{O}{3}] LFs at similar redshifts. \cite{Champagne2024} observed suppression of [\ion{O}{3}] LFs at $L_{\rm [O\,III]} < 10^{42.6}$ \si{erg.s^{-1}} in protoclusters at $z = 5.4$, 6.1, and 6.6, in contrast to the steep power-law shape of the field [\ion{O}{3}] LFs. Notably, only the $z=6.6$ protocluster in \cite{Champagne2024} is associated with quasars, whereas the $z=5.4$ and $z=6.1$ protoclusters are not. This implies the suppressed faint-end [\ion{O}{3}] LF in protoclusters is not driven by quasar feedback.
 
The flatter faint-end slopes in protoclusters at $z > 5$, which have values similar to lower redshifts, suggest accelerated evolution and more evolved stellar populations in these overdense regions. Recent JWST studies of more high-redshift protoclusters further support evidence of accelerated star formation and the presence of evolved stars \citep[e.g.,][]{Helton2024a,Helton2024}. \cite{Morishita2024} found a higher fraction of low-equivalent-width emitters with strong Balmer breaks in overdensities at $z = 5.7$ and 5.8. These galaxies may have recently experienced declines/lulls in star formation after bursts. \cite{Witten2025} reported a galaxy in a $z = 7.9$ protocluster \citep{Morishita2023, Hashimoto2023} with a prominent Balmer break and a lower [\ion{O}{3}]/[\ion{O}{2}] ratio than most galaxies at similar redshifts, revealing mature stellar populations of $\sim 80$ Myr.

On the other hand, low-mass field galaxies at $z>5$ may undergo active star formation, for example recurring bursts \citep{Tacchella2016, McClymont2025}, along with rapid mass assembly over approximately 300 Myr.  These processes may help shape the LF and ultimately establish its form by $z \sim 4$. This scenario can be further investigated by examining galaxy properties such as the Balmer break strength and fraction across different environments and luminosity/mass ranges \citep{Endsley2024, Endsley2024b, Dome2024, Gelli2025}. Future observations with larger, independent survey areas and greater depth (e.g., the ongoing Cycle-3 pure-parallel grism program SAPPHIRES; \citealt{sapphires_edr}) are needed to mitigate cosmic variance, capture more faint-end samples, and systematically investigate the redshift evolution of LFs and the environmental effects.

Our results have revealed the impact of environments on galaxy evolution at $z>5$. The distinct shapes of LFs in different environments, as shown in Figure \ref{fig:HaLF}, are clearly visible regardless of the exact value of $\alpha$. We caution that our LF bins with high completeness (completeness $>0.7$) have not yet reached the faintest regime \citep[e.g., $L\lesssim0.01L_*$;][]{Bouwens2015, Bouwens2021}. \citet{Fu2025} obtained an H$\alpha$ LF at $z \approx 4.5$ from the JWST MAGNIF survey (Cycle-2 GO-2883; PI: Sun, F.) in four lensing cluster fields down to $L_{\rm H\alpha} = 10^{40.3}$ \si{erg\,s^{-1}}\, $\gtrsim10\times$ deeper than our survey in blank field. \citet{Fu2025} measured a faint-end slope of $\alpha \approx -1.8$, which is much steeper than $\alpha$ measured from this study. 
Cosmic variance across different fields may contribute to these differences, highlighting the need for larger surveys to mitigate these effects. Although studies of high-redshift emitters in lensing cluster fields may provide stronger constraints on the faint-end slope of field galaxies,  the limited survey volume of these studies makes it challenging to identify large protoclusters like those highlighted in this work.   Future observations of faint galaxies in protoclusters are needed to more accurately constrain the faint-end LF shapes of protocluster galaxies and their evolution across cosmic time.

\bigskip

\section{Clustering Analysis}\label{sec:clustering}

While luminosity functions provide insights into the overall properties and evolution of the HAE population, clustering analysis reveals the growth of large-scale structures and their dark matter halo properties. In this section, we explore the clustering properties of HAEs at $z=4-6$ through their two-point auto-correlation functions. The clustering studies are conducted in the redshift range of $z = 4 - 5$ and $z = 5 - 6$, which differs from those used in the \ha\ LF part. This configuration ensures the same $\Delta z$ for both bins and keeps the volume size along the LOS not too large, thus minimizing the impact of evolution.

\subsection{The random galaxy catalog}

To simulate the spatial distribution of the galaxy without any clustering signal, we generate random galaxy catalogs at $z=4-6$ based on the measured \ha\ LFs. For each of the two redshift bins, we randomly sample a Schechter function from the posterior distribution of the MCMC-derived \ha\ LF models. We estimate the expected galaxy counts within a box spanning the redshift bin, which fully covers the WFSS footprint with a projected area of 20$\times$20 cMpc$^2$. We draw random positions within the box, totaling 100 times the expected galaxy number, and assign line fluxes according to the selected Schechter function.  For each random galaxy, we estimate completeness based on its \ha\ line flux and the RMS value at its \ha\ line position, as discussed in Appendix \ref{sec:completeness_function}.  The completeness of galaxies outside the coverage of WFSS equals zero.   We determine if a random galaxy is observable by sampling from a binomial distribution, with the completeness value as the probability of detection. We repeat the experiment 100 times, resulting in 100 random galaxy catalogs. Each random galaxy catalog has a galaxy number density 100 times greater than that of the observed catalogs to minimize the Poisson noise from the random sources. These catalogs contain no clustering signals while preserving the same selection function as the WFSS observations.

\subsection{The auto-correlations of \ha\ emitters at $4< z<6$}\label{sec:hae_acf}

\begin{figure*}[htbp!]
    \centering
    \includegraphics[width=\linewidth]{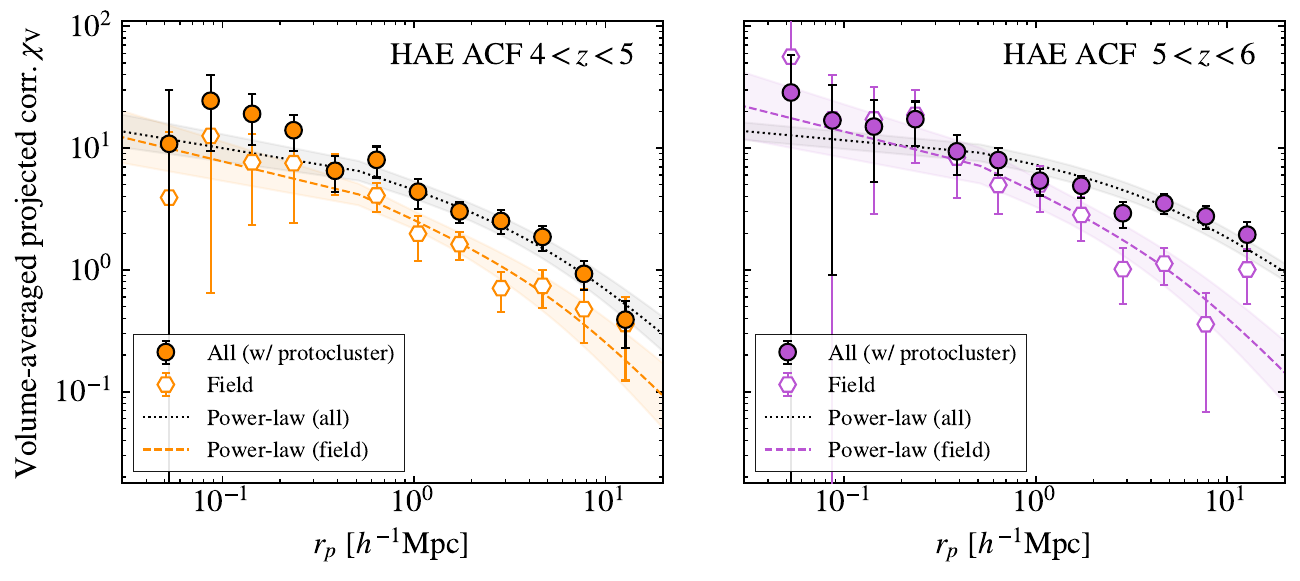}
    \caption{The volume-averaged projected auto-correlation functions of HAEs. \textit{Left:} The filled orange dots show the auto-correlation function of all HAEs at $4\lesssim z<5$, including the protocluster at $z\approx 4.41$. Its best-fit power-law model and uncertainty are shown as the black dotted line and gray-shaded region. The open orange hexagons represent the auto-correlation function that excludes the $z\approx 4.41$ 
    protocluster. Its best-fit power-law model and uncertainty are shown as the dashed orange line and orange-shaded region. \textit{Right:} The filled purple dots show the auto-correlation function of all HAEs at $5\lesssim z<6$, including the protocluster at $z\approx5.19$. Its best-fit power-law model and uncertainty are shown as the black dotted line and gray-shaded region. The open purple hexagons represent the auto-correlation function that excludes the $z\approx 4.41$ 
    protocluster. Its best-fit power-law model and uncertainty are shown as the dashed purple line and purple-shaded region.   }
    \label{fig:HAE_ACF}
\end{figure*}

In the clustering analysis, we include only HAEs and random galaxies at $z=4-6$ with $L_{\rm H\alpha} > 10^{41.5}~{\rm erg~s^{-1}}$, corresponding to a SFR limit of $1.41~M_\odot/{\rm yr}$. This aims to exclude the faintest \ha\ luminosity bins in the \ha\ LFs (Figure \ref{fig:HaLF}), where the completeness correction is subject to large systematic uncertainties. The luminosity cut yields a total of 863 HAEs. We group galaxies with projected separations smaller than 10 physical kpc and LOS separations smaller than 500 \si{km\,s^{-1}} into a system, assuming they are interacting and gravitationally bound. This separation corresponds to approximately 1.5\arcsec\ at $z \approx 4 - 6$, consistent with the definition of a galaxy system that has been applied to $z \sim 6$ [\ion{O}{3}] emitters \citep{Matthee2023, Eilers2024} and well within the defining separation for mergers \citep[e.g.,][]{Puskas2025}. About 17\% of the HAEs have companion galaxies and are thus grouped into systems. We finally obtain \NHAEcl\ systems at $4<z<6$ in total. \xj{The grouped systems are used to compute both galaxy–galaxy and galaxy–random pair counts below.}
 
We calculate the galaxy auto-correlation function using the Landy-Szalay (LS) estimator \citep{LS}:

\begin{equation}
\xi_{gg}(\vec{r})=\frac{N_{\mathrm{r}}\left(N_{\mathrm{r}}-1\right)}{N_{\mathrm{g}}\left(N_{\mathrm{g}}-1\right)} \frac{G G (\vec{r})}{R R(\vec{r})}-\frac{N_{\mathrm{r}}-1}{N_{\mathrm{g}}} \frac{G R(\vec{r})}{R R(\vec{r})}+1,
\end{equation}
where $\vec{r}$ is the separation vector, which can be projected onto the LOS direction as the LOS distance $r_\pi$ and onto the plane of the sky as the projected distance $r_p$.  $G G (\vec{r})$ and  $R R(\vec{r})$ are the number of observed HAE pairs and random galaxy pairs with separation $\vec{r}$, respectively. $G R(\vec{r})$ is the number of HAE-random pairs. $N_{\rm g}$ and $N_{\rm r}$ are the size of the HAE and random galaxy catalogs.  We use \textsc{Corrfunc}\footnote{\url{https://github.com/manodeep/Corrfunc}} \citep{Sinha2019, Sinha2020} to calculate the pair counts.

We define the volume-averaged projected correlation function $\chi_V$ by integrating $\xi_{gg}(\vec{r})$ over a cylindrical shell \citep{Hennawi2006}:

\begin{equation}\label{eq:chiV}
\chi_V\left(r_p\right)=\frac{2}{V} \int_{r_{p, \min }}^{r_{p, \max }} \int_0^{r_{\pi,\max }} \xi\left(r_p, r_\pi\right) 2 \pi r_p \mathrm{~d} r_p \mathrm{~d} r_\pi,
\end{equation}
where $r_{\pi, \max}$ is the integration limit along the LOS direction. We set $r_{\pi, \max}$ to be 8 $h^{-1}$cMpc, corresponding to $\sim 1000$ km s$^{-1}$ at $z\approx 4-6$. $V$ is the volume of the cylindrical shell in the $r_p$ bin ($r_{p, \min }<r_p<r_{p, \max }$), which can be expressed as  $V=\pi \left(r_{p, \max }^2-r_{p, \min }^2\right) \cdot r_{\pi, \max}$.  We compute $\chi_V$ for the HAE sample at $z=4-5$ and $z=5-6$ separately.  We construct the covariance matrix by bootstrapping, in which we sample the new HAE catalog with replacement each time and calculate the auto-correlation function using different random catalog realizations. The uncertainties are derived from the diagonal of the matrix.  

To account for the effect of the protoclusters at $z = 4.41$ and $z = 5.19$, we also compute the correlation functions for the field galaxies. We run the Friends-of-Friends algorithm on the new catalog, which has had the luminosity and grouping cuts applied. The two largest structures of the protoclusters are within $z = 4.39 - 4.43$ and $z = 5.16 - 5.20$, with 70 systems in $z = 4.39 - 4.43$ and 89 in $z = 5.16 - 5.20$. We therefore compute $\chi_V$ for field galaxies by masking HAEs and random galaxies within the redshift ranges $z = 4.39 - 4.43$ and $z = 5.16 - 5.20$. We only mask the most extreme structures in the two protoclusters, rather than the entire redshift range that the protocluster occupies. Our goal is to explore the impact of protocluster geometry while preserving a sufficient field sample for the correlation calculation.  The volume-averaged projected correlation function $\chi_V$ with and without protoclusters are shown in Figure \ref{fig:HAE_ACF}.  We
parameterize the auto-correlation functions by a power-law:
\begin{equation}
\xi(r)=\left(r / r_0\right)^{-\gamma},
\end{equation}
where $r_0$ represents the characteristic scale length where $\xi(r)=1$. The auto-correlation function measurements are shown in Figure \ref{fig:HAE_ACF}. We run MCMC to fit the measured $\chi_V$, and the best-fit $r_0$ and $\gamma$ for scenarios with and without protoclusters are listed in Table \ref{tab:acf_hm_sm}.

When considering field galaxies alone, the auto-correlation function follows a power-law shape with $r_0=4.61^{+1.00}_{-0.68}$ $h^{-1}$cMpc and $\gamma=1.59^{+0.23}_{-0.25}$ at $4<z<5$,  and $r_0=6.23^{+1.68}_{-1.13}$ $h^{-1}$cMpc and $\gamma=1.63^{+0.24}_{-0.26}$ at $5<z<6$.  However, when protoclusters are included in the clustering analysis, the auto-correlation function becomes flattened, with a smaller $\gamma$ and larger $r_0$. At $4 < z < 5$, it gives $r_0 = 8.29^{+1.66}_{-1.21}$ $h^{-1}$cMpc and $\gamma = 1.34^{+0.13}_{-0.14}$, and at $5 < z < 6$, $r_0 = 19.71^{+3.38}_{-2.90}$ $h^{-1}$cMpc and $\gamma = 1.02^{+0.03}_{-0.02}$.

\begin{table*}
\centering
		\begin{tabular}{cccccccc}
        \hline
        \hline
			redshift range & scenario &  $r_0/h^{-1}{\rm Mpc}$ & $\gamma$ & $\log (M_{ h, \rm UM}/M_\odot)$ & $b$ & $\log (M_{ h, b}/M_\odot)$  \\
			\hline
			\multirow{2}{*}{$4 < z < 5$} &  w/ protocluster &  $8.29_{-1.21}^{+1.66}$ & $1.34_{-0.14}^{+0.13}$ & $11.21_{-0.33}^{+0.36}$ & -- & -- \\
			& field &$4.61_{-0.68}^{+1.00}$ & $1.59_{-0.25}^{+0.23}$ & $11.21_{-0.32}^{+0.35}$  & $4.11\pm0.12$ & $11.20\pm0.05$ \\
            \hline
			\multirow{2}{*}{$5 < z < 6$} & w/ protocluster & $19.71_{-2.90}^{+3.38}$ & $1.02_{-0.02}^{+0.03}$ & $11.06_{-0.32}^{+0.36}$ & -- & --  \\
			& field & $6.23_{-1.13}^{+1.68}$ & $1.63_{-0.26}^{+0.24}$ & $11.04_{-0.32}^{+0.34}$  & $5.90\pm0.08$ & $11.25\pm0.02$  \\
        \hline
		\end{tabular}
	\caption{HAE auto-correlation function and the estimated halo mass ($M_{ h, \rm UM}$) 
    of the galaxies from the \textsc{UniverseMachine}. As a complementary approach, we also report the bias ($b$) and corresponding halo masses ($M_{h,b}$) for the field sample. We do not measure the bias for the full sample (w/ protocluster) because the shape of the auto-correlation function is distorted by the protocluster’s geometry.}
	\label{tab:acf_hm_sm}
\end{table*}

\subsection{The impact of protocluster geometry to the auto-correlation functions}\label{sec:acf_filament}

 \begin{figure*}[htbp!]
    \centering
    \includegraphics[width=0.495\linewidth]
    {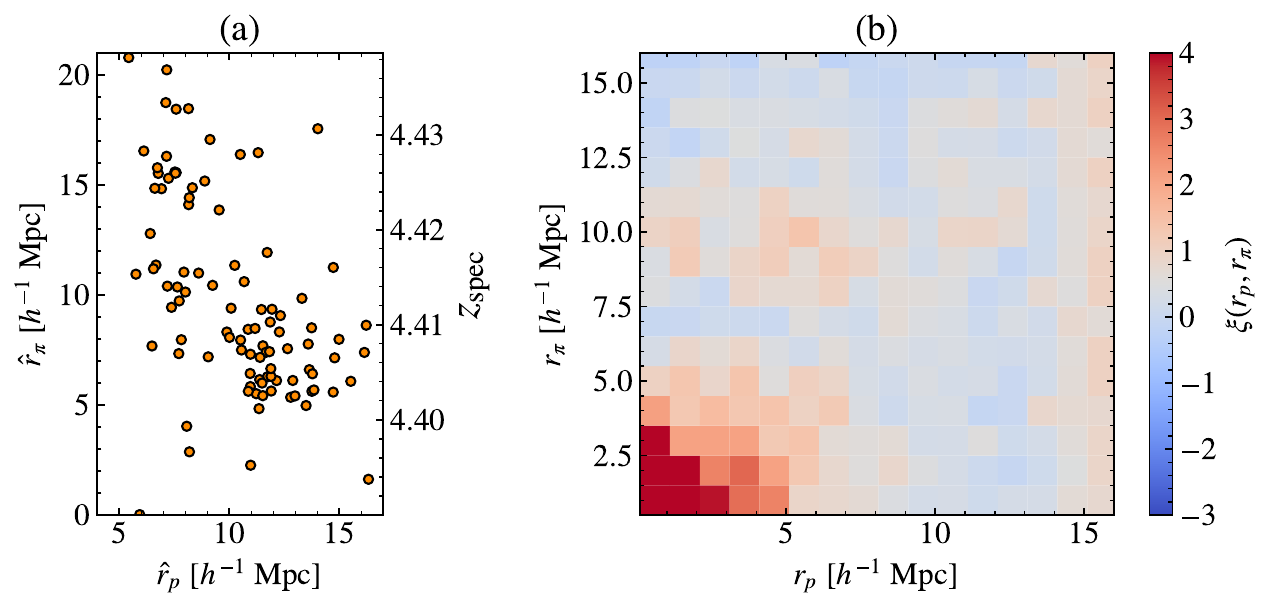}
    \includegraphics[width=0.495\linewidth]{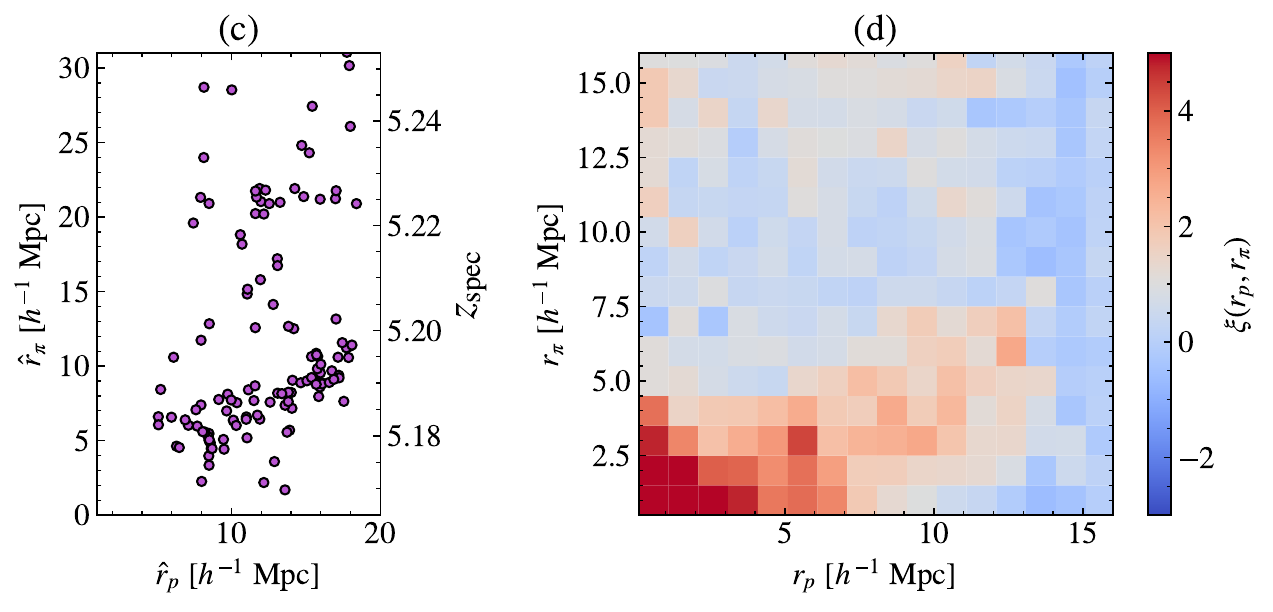}
    \caption{(a): The projected structure of the $z\approx4.41$ protocluster.  $\hat{r}_p$ and $\hat{r}_\pi$ are coordinates of galaxies with respect to (RA, Dec, $z$) = (189.0159, 62.1718, 4.39). (b):  HAE auto-correlation at $4< z<5$ as a function the projected distance of $r_p$ and LoS distance $r_\pi$.  (c): The projected structure of the $z\approx5.19$ protocluster.  $\hat{ r}_p$ and $\hat{r}_\pi$ are coordinates of galaxies with respect to (RA, Dec, $z$) = (189.0275, 62.1602, 5.165). (d):  HAE auto-correlation at $5 < z<6$ as a function of the projected distance $r_p$ and LOS distance $r_\pi$.    The filamentary structures of protoclusters create elongated patterns in the 2D $\xi(r_p, r_\pi)$ planes, which result in flattened power-law slopes of $\chi_V$ when collapsing over $r_\pi = 0-8~h^{-1}\mathrm{cMpc}$.  }
    \label{fig:filamentary_xi}
\end{figure*}

As shown in Figure \ref{fig:HAE_ACF}, when including the protoclusters,  the auto-correlation function slope $\gamma$ deviates from the typical range of $ 1.6 - 2.0$ reported in the literature \citep[e.g.,][]{Hennawi2006, Geach2012, Eilers2024}. 

We analyze the 2D auto-correlation functions  $\xi(r_p, r_\pi)$  in comparison with the protocluster geometry in Figure \ref{fig:filamentary_xi}.  The protoclusters at $z=4.41$, as illustrated in the panel (a) of Figure \ref{fig:filamentary_xi}, span $10~h^{-1}\mathrm{cMpc}$ along $r_p$ and $15~h^{-1}\mathrm{cMpc}$ along $r_\pi$. It exhibits elongation along the $r_\pi$ direction compared to a spherical geometry. As a result, the $\xi(r_p, r_\pi)$ signal shows an excess along $r_p$, deviating from the typical two-dimensional clustering pattern (see panel (b) of Figure \ref{fig:filamentary_xi}). The elongated excess is particularly prominent at $r_\pi \gtrsim 5~h^{-1}\mathrm{cMpc}$, extending to $r_p \approx 8~h^{-1}\mathrm{cMpc}$. This leads to a flattened power-law slope of $\chi_V$, which is computed by collapsing along $r_\pi$ over the range $0-8~h^{-1}\mathrm{cMpc}$ (Equation \ref{eq:chiV}). For the protocluster at $z=5.19$, the prominent filament extends to about $15~h^{-1}\mathrm{cMpc}$ along $r_p$ and $10~h^{-1}\mathrm{cMpc}$ along $r_\pi$. As shown in the panel (c) of Figure \ref{fig:filamentary_xi}, it exhibits a distinct orientation in the projected $r_p$-$r_\pi$ plane, from $(\hat{r}_p, \hat{r}_\pi)\approx(7,5) ~h^{-1}{\rm cMpc}$ to $(\hat{r}_p, \hat{r}_\pi)\approx(20,10) ~h^{-1}{\rm cMpc}$. The galaxy pairs within the filament lead to similar elongation in the $\xi(r_p, r_\pi)$ plane, spanning approximately from $(r_p, r_\pi) = (5, 2.5)~h^{-1}{\rm cMpc}$ to $(r_p, r_\pi) = (13, 7.5)~h^{-1}{\rm cMpc}$. This pattern naturally results in an increased amplitude of  $\chi_V$ at $r_p > 5~h^{-1}{\rm cMpc}$ when averaging over $r_\pi$, and thus the flattened power-law slope.

The analysis above cautions that galaxy clustering within a limited FoV, especially in the presence of protoclusters, does not necessarily follow a power-law with standard slopes around $1.6-2.0$. This situation is more common in the high-redshift Universe, where overdensities and filamentary structures have recently been found to be ubiquitous \citep[e.g.,][]{Helton2024, Li2024, Morishita2024}.

\subsection{The implication for the dark matter halo masses of HAEs.}\label{sec:um}

\begin{figure*}
    \centering
    \includegraphics[width=\linewidth]{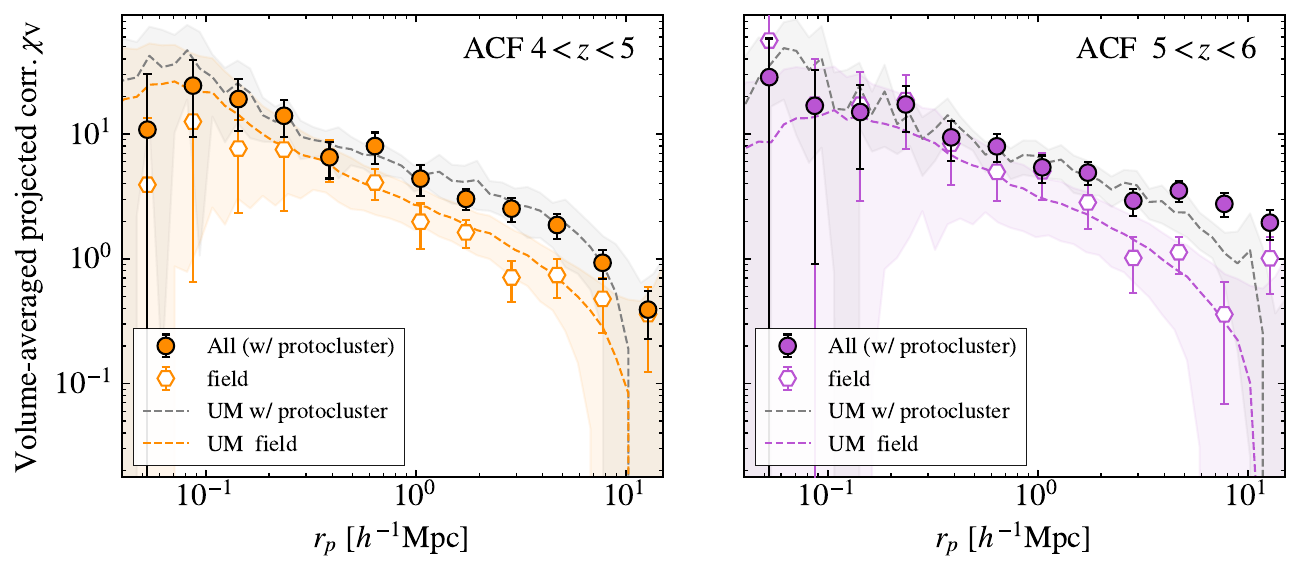}
    \caption{The measured HAE auto-correlation functions and models from the UniverseMachine. \textit{Left}: The measured HAE auto-correlation function of all HAEs at $4< z < 5$, including the $z=4.41$ protoclusters, is shown as filled orange dots, and that of the field HAEs is shown as orange-edged white hexagons. The median \textsc{UniverseMachine} galaxy auto-correlation function at $4< z < 5$ dominated by protoclusters is shown as the dashed gray lines, with the gray-shaded region indicative of the $1\sigma$ range.  Likewise, the \textsc{UniverseMachine} auto-correlation functions for field galaxies are represented by the dashed orange line and the orange-shaded region. \textit{Right:} Similar to the left panel. The measured auto-correlations of all HAEs are shown as filled purple dots and purple-edged white hexagons, respectively. The median \textsc{UniverseMachine} galaxy auto-correlation functions dominated by protocluster and field galaxies within the footprint are dashed gray and purple lines. As shown in both panels, the \textsc{UniverseMachine} results and our measured auto-correlation function are in good agreement.}
    \label{fig:UM_gal_ACF}
\end{figure*}

\begin{figure}
    \centering
    \includegraphics[width=1\linewidth]{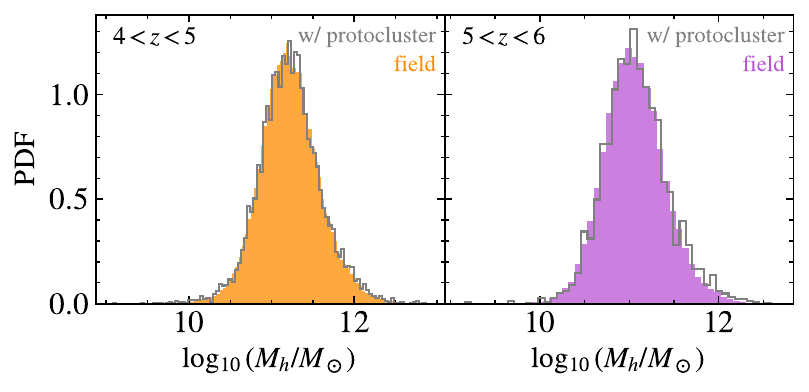}
    \caption{The halo mass ($M_h$) distribution of (sub-)halos in the UniverseMachine mock surveys. \textit{Left}: The black and orange histograms denote the $M_h$ distribution of all HAEs including the protoclusters (\texttt{w/ protoclusters}) and HAEs in the fields  (\texttt{field}), which correspond to the models in the left panel of Figure \ref{fig:UM_gal_ACF}. \textit{Right}: Similar to the left panel. The black and purple histograms correspond to the models in the right panel of Figure \ref{fig:UM_gal_ACF}.}
    \label{fig:UM_Mh_distribution}
\end{figure}

 \begin{figure*}
    \includegraphics[width=\textwidth]{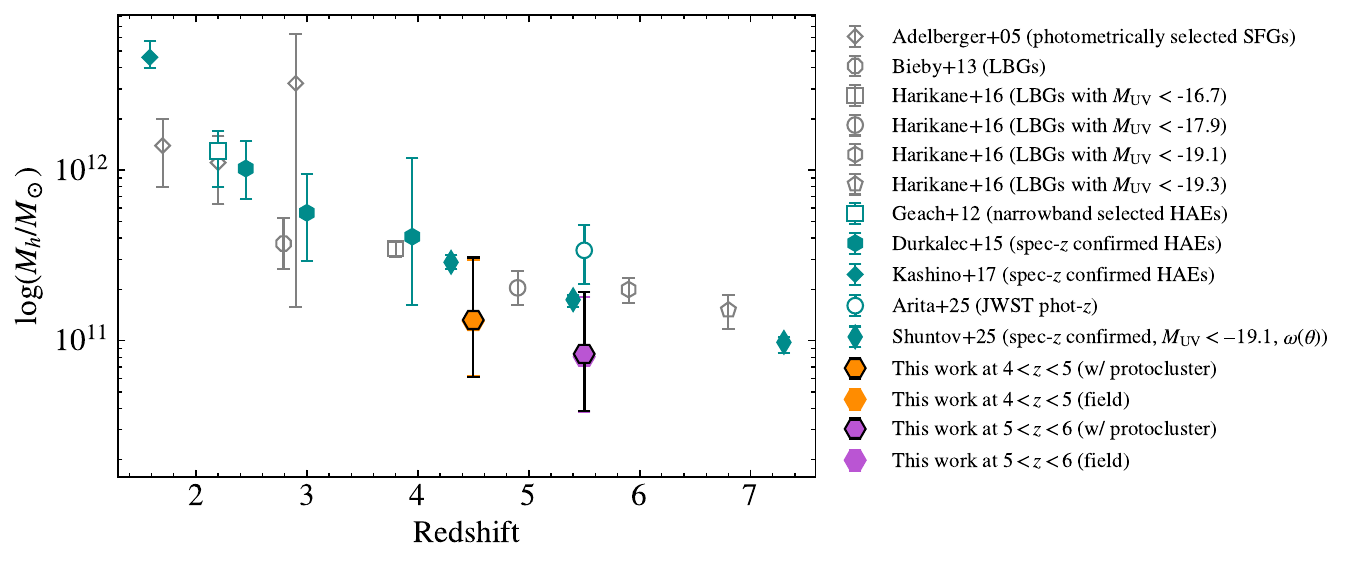}
    \caption{Halo masses of star-forming galaxies as a function of redshift. The literature data points include color-selected star-forming galaxies from \cite{Adelberger2005}, Lyman-Break galaxies from \cite{Bielby2013} and \cite{Harikane2016}, narrowband-selected HAEs from \cite{Geach2012}, spectroscopically confirmed HAEs from \cite{Durkalec2015, Kashino2017},  JWST phot-$z$ selected galaxies from \cite{Arita2025}, and grism-selected emitters from \cite{Shuntov2025} with halo masses estimated from the angular correlation function ($\omega(\theta)$).}
    \label{fig:hm_z}
\end{figure*}

We use the \textsc{UniverseMachine} simulations \citep{Behroozi2019} to reproduce the observed correlation functions and estimate the dark matter halo mass distribution.  The \textsc{UniverseMachine} generates mock universes by applying empirically derived relations from observations to dark matter halo merger trees. We refer to \cite{Behroozi2019} for more details on the implementation of the simulation and its predictions for high-redshift observations.  We start from the Small MultiDark-Planck (SMDPL) simulations in the MultiDark and Bolshoi Project\footnote{\url{https://www.cosmosim.org/}} \citep{Klypin2016}, which have box sizes of 400 $h^{-1}$cMpc.  SMDPL simulations have box sizes about 25 times larger than the WFSS FoV while preserving adequate dark matter particle resolution ($10^8\, h^{-1}M_\odot$ per dark matter particle). Such large box sizes can minimize the edge effect in clustering analysis, which arises from counting galaxies multiple times due to periodic boundary conditions.  We use the public \textsc{UniverseMachine} DR1 code\footnote{\url{https://bitbucket.org/pbehroozi/universemachine}} to generate 50 lightcone realizations with areas of $32\times32$ arcmin$^{2}$ over $3<z<7$. Each lightcone realization spans a comoving volume of $10^{6.5}$\,Mpc$^3$ across $z \approx 4-5$ and $10^{6.4}$\,Mpc$^3$ at $z = 5-6$, about 15 times larger than our grism survey volume. Although the \textsc{UniverseMachine} lightcone products provide predictions for galaxy stellar masses ($M_*$), specific star formation rates (sSFR), and star formation rates (SFR), these properties are extrapolated from pre-JWST observations and may not fully align with recent JWST findings. Recent JWST observations have revealed more bursty star formation in high-redshift galaxies \citep[e.g.,][]{Sun2023, Endsley2024, Boyett2024}, necessitating updates to previous stellar population models.  The \textsc{UniverseMachine} output SFRs may underestimate instantaneous star formation intensity in high-redshift galaxies, leading to lower \ha\ luminosities when directly converted from the SFRs. Additionally, the SFRs from the \textsc{UniverseMachine} lightcone products are averaged over the past 20 Myr, while \ha-derived SFRs are more sensitive to star formation within the most recent 10 Myr \citep{Kennicutt2012}. Therefore, we re-calibrate the mock SFRs by adding an offset that is parameterized using the mock $M_*$ and sSFRs. The offset is determined by aligning the mock \ha\ LF and SFR-$M_*$ distribution with the observed \ha\ LF and \ha-derived SFR-$M_*$ distribution, where $M_*$ is obtained from SED modeling.  We provide a full description of the calibration and conversion into the observed \ha\ line flux in Appendix \ref{sec:um_calibration}. \xj{The re-calibration only modifies the observables from the simulation, while keeping the original stellar mass–halo mass relation, and thus the galaxy–halo connection in terms of $M_*$. It ensures that the selection function is applied in the same manner as in real grism surveys, as described below.}

We randomly select 40 rectangular regions of $8\times 8$ arcmin$^2$ at $z\approx4-5$ and $z\approx5-6$ in each lightcone realization. This results in a total of 2,000 mock surveys for each redshift bin. The mock survey FoVs are set to match our observations, to account for sample variance in such a small area. We aim to reproduce both the clustering dominated by protocluster structures and the clustering with field galaxies. We select galaxies with observed \ha\ luminosities greater than $10^{41.5} ~\mathrm{erg} \, \mathrm{s}^{-1}$, which are derived by applying the dust attenuation $A_{\rm UV}$ in the mock catalog assuming the SMC attenuation law \citep{Pei1992}. The luminosity cut is the same as that in our clustering measurements.  We run a Friends-of-Friends algorithm to identify protoclusters where member galaxies have projected separation smaller than 500 physical kpc and LOS separation smaller than 500 km s$^{-1}$ (same as that in Section~\ref{sec:hae_acf}). At $4 < z < 5$ ($5 < z < 6$), if the identified protocluster contains more than 20\% (30\%) of the galaxies within the entire redshift bin, we label this case as \texttt{w/ protocluster}. In this case, we calculate the galaxy auto-correlation using all the galaxies within the FoV, \xj{including both protocluster and field galaxies over the full redshift bin.} The minimum galaxy fraction within protoclusters ensures that these structures have a measurable impact on the observed clustering signal shapes. We calculate the auto-correlation of field galaxies by masking any structures that contain more member galaxies than those in the $z=4.41$ and $z=5.19$ protoclusters. 
It ensures that the environments of involved galaxies in this situation are comparable to those of our observed HAEs in the fields. We label this case as \texttt{field}.  \xj{All analyses in the mock surveys mirror the measurements in the real observations.} 
 Among the  2,000 mock surveys, we find that only 30 instances at $z=4-5$ and 21 instances at $z=5-6$ exhibit structures that meet our criteria for significant protoclusters.
The rarity of protoclusters in simulations contrasts with their ubiquity in recent JWST observations.  We compare and discuss the abundance of protoclusters in observations and simulations in \S\ref{sec:um_od_num}.

We show the clustering results of \textsc{UniverseMachine} mock surveys in Figure \ref{fig:UM_gal_ACF}. The variance from these mock surveys (i.e., the shaded region in Figure \ref{fig:UM_gal_ACF}) represents uncertainties arising from cosmic and sample variance.  Both the mock clustering in the \texttt{w/ protocluster} and \texttt{field} scenarios are in good agreement with the observed HAE auto-correlations within 1$\sigma$. For the galaxy auto-correlation in the \texttt{w/ protocluster} scenario, the amplitude is elevated at $\gtrsim 1h^{-1}$ cMpc. \xj{The \texttt{field} case changes by 0.6 (0.7) dex from $r=1$ to $5h^{-1}{\rm Mpc}$ at $z=4-5$ ($z=5-6$), while the \texttt{w/ protocluster} scenario changes by only 0.4 (0.5) dex. These results indicate that, on average, the power-law slope is flatter at large scales when protoclusters are included, despite the large variance.} The consistency between simulations and observations confirms that the geometry of filamentary structures drives the flat auto-correlation shape in an overdense small region.  \xj{We note that there are no free parameters when extracting clustering signals from the recalibrated \textsc{universemachine}; only the selection function is applied. The agreement between the observed and simulated clustering suggests that the selected halos can reasonably represent our observed galaxies.}

The distribution of dark matter (sub-)halo masses ($M_h$) of galaxies in the mock surveys is shown in Figure \ref{fig:UM_Mh_distribution} and summarized in Table \ref{tab:acf_hm_sm}.  The HAEs at $4 < z < 5$ are hosted by (sub-)halos with a typical $M_h$ around $10^{11.2} M_\odot$, and at $5 < z < 6$, the typical $M_h$ is about $10^{11.1} M_\odot$. The $1\sigma$ uncertainties are 0.4 dex, which have incorporated systematics introduced by cosmic and sample variance.

\xj{As a complementary approach, we estimate the bias of HAEs and derive the corresponding halo masses. We restrict the measurements to the field HAE sample to avoid distortions from the protocluster. The biases ($b$) are obtained by fitting the measured auto-correlation functions at $r_p > 1\,h^{-1}{\rm Mpc}$ to that of the underlying dark matter field, adopting the \textsc{camb} transfer function \citep{Lewis2011} and the bias model of \cite{Tinker2010}. The results are listed in Table~\ref{tab:acf_hm_sm}. The halo masses of field HAEs inferred from \textsc{UniverseMachine} ($M_h$) and from the bias measurements ($M_{h,b}$) are in good agreement within $1\sigma$. We adopt the \textsc{UniverseMachine} values as our fiducial estimates, as they more robustly account for the sample and cosmic variance given the limitations of our survey volume. Using the bias-derived values does not alter our conclusions.}

We place our derived $M_h$ for HAEs in the context of star-forming galaxies from the literature in Figure \ref{fig:hm_z}. The literature values of $M_h$ are all based on clustering analyses, though the methods used to derive $M_h$ vary, e.g., halo occupation distribution models \citep{Seljak2000}. 
We caution that different works use varying selection criteria and luminosity cuts for galaxies, whereas the derived halo masses strongly depend on galaxy properties. Our sample size and survey volume are not sufficient for a luminosity-dependent clustering analysis. Future surveys, such as COSMOS-3D (GO-5893, PI: Kakiichi), will provide further insights into dark matter halos and their dependence on galaxy properties.

\subsection{The rarity of significant protoclusters in simulations}\label{sec:um_od_num}

As mentioned in \S\ref{sec:um}, we conduct a systematic search in \textsc{UniverseMachine} for protoclusters similar to the observed protoclusters at $z\approx 4.41$ and $z\approx 5.19$. We require at least $20\%$ or 30\% of the mock galaxies with $L_{\rm H\alpha}>10^{41.5}\,{\rm erg\,s^{-1}}$ at $z\approx4-5$ and $z\approx5-6$ are members of the protoclusters. Across 2,000 mock surveys spanning 50 \textsc{UniverseMachine} lightcone realizations, we identify structures similarly significant to the observed ones in 30 instances at $z=4-5$ and 21 instances at $z=5-6$. By tracing these structures back to their positions, we find they correspond to nine unique protoclusters at $z=4-5$ and four at $z=5-6$ in the lightcones. These protoclusters are repeatedly identified in the mock surveys, as they are (partially) covered by the random FOVs placed on the lightcones. It suggests that the probability of identifying a structure comparable to our observed protoclusters in \textsc{UniverseMachine} is only 1.5\% at $4<z<5$ and 1.1\% at $5<z<6$. 

In comparison, the effective survey volume of our grism observations (see \S\ref{sec:HaLF} and Appendix \ref{sec:Appendix_MLE_result}) is approximately $199,824\,\text{Mpc}^3$ at $z=3.75-5$ and $151,163\,\text{Mpc}^3$ at $z=5-6$, within which we identify one significant protocluster in each redshift range. It implies a number density of such structure of about $10^{-5.2}$\,Mpc$^{-3}$ for each redshift bin.  Based on the number density estimated from our grism observations, we would expect to detect at least one prominent protocluster in each $8\times8$\,arcmin$^2$ mock survey, about 15–20 in each $32\times32$\,arcmin$^2$ lightcone realization, and $900-1200$ similar structures across the entire $(400\,h^{-1}\,\text{Mpc})^3$ \textsc{UniverseMachine} box. The contrast reveals the rarity of such significant protoclusters in \textsc{UniverseMachine}.  We test the impact of the calibration parameters by loosening their range (see Appendix \ref{sec:um_calibration}) and find that the number of protoclusters varies from 1 to 14 at $4<z<5$ and from 1 to 11 at $5<z<6$. The probability of finding prominent protoclusters ranges from 0.05\% to 1.9\% at $4<z<5$ and from 0.05\% to 2.4\% at $5<z<6$. We thus qualitatively conclude that the probability of finding prominent protoclusters in \textsc{UniverseMachine} is low, on the order of $\lesssim 1\%$.

In fact, the discovery of these prominent protoclusters in our grism surveys is not merely a coincidence. The ubiquity of overdensities has been revealed by multiple surveys and independent sightlines \citep{Kashino2023, Wang2023, Champagne2024, Helton2024a, Helton2024, Morishita2024, sapphires_edr, sapphires_od}. 
Overdensities are commonly detected within areas spanning one to two JWST pointings, though the size and scale of these structures should be evaluated on a case-by-case basis.  We caution that literature studies employ diverse methods to characterize the significance of protoclusters, often using different apertures to calculate the overdensity values. The conclusions of this paper are limited to prominent structures that contain more than 20\% of galaxies in specific redshift bins. We leave it to future studies to conduct a more quantitative assessment of protocluster abundance in both simulations and observations.  This will require a standardized criterion for defining the significance of protoclusters across different surveys and simulations, as well as improved calibration of simulations and a more rigorous treatment of selection effects.

\bigskip 

\section{Summary}

In this paper, we present a comprehensive study of the luminosity functions (LFs) and clustering properties of \ha\ emitters (HAEs) in the GOODS-North field. To spectroscopically identify these emitters across different redshifts, we develop a semi-automated algorithm using JWST/NIRCam WFSS data. By combining the F356W grism data from the CONGRESS program with the F444W grism data from the FRESCO program, we identify \NHAE\ \ha\ emitters in the redshift range $3.75 < z < 6$. Among these, we detect two significant protoclusters at $z = 4.41$ and $z = 5.19$, containing 98 and  144 member galaxies, respectively. We measure the \ha\ LFs for galaxies in both the protoclusters and the fields. We perform a SFR-limited clustering analysis with HAE auto-correlation functions. Our main conclusions are as follows:

\begin{itemize}
    \item We measure the \ha\ LFs in two redshift bins, $3.75 < z < 5$ and $5 < z < 6$, down to a luminosity limit of $10^{41.5}$ erg s$^{-1}$. We also calculate the \ha\ LFs for the $z = 4.41$ and $z = 5.19$ protoclusters, as well as for field galaxies. In each redshift bin, the overall \ha\ LF, protocluster \ha\ LF and the field LF show similar characteristic luminosities. At $3.75 < z < 5$, the \ha\ LFs for all HAEs, protocluster galaxies, and field galaxies share similar faint-end slopes ($\alpha \approx -1.3$). In contrast, at $5 < z < 6$, the field \ha\ LF shows a steeper slope of $-1.8$, while the protocluster \ha\ LF maintains a slope consistent with the $z < 5$ values within $1\sigma$. 
    We conclude that at $z>5$, galaxies in protoclusters are more evolved, with their LF matched that at $z<5$. In contrast, the LF of field galaxies are under rapid development, which flattens at $z\sim4$.

    \item The \ha-based cosmic star formation rate density (uncorrected for dust) at $3.75 < z < 5$ and $5 < z < 6$  is 0.27 dex and 0.47 dex higher than the \cite{Madau2014} model, respectively. Across $3.75 < z < 5$, over $25^{+6}_{-3} \%$ of cosmic star formation occurs in the protocluster,  generally consistent with the simulation prediction \citep{Chiang2017, Lim2024}. At $5 < z < 6$, the protocluster contributes over $55 ^{+16}_{-11}\%$ of the cosmic star formation. These fractions may be subject to cosmic variance due to the limited survey volume.

    \item We conduct a three-dimensional clustering analysis of HAEs with $L_{\rm H\alpha}>10^{41.5}$\,\si{erg\,s^{-1}}, corresponding to ${\rm SFR} > 1.41$\,\si{M_\odot\,yr^{-1}}. The auto-correlation function of field HAEs at $4 < z < 5$ follows a power-law shape with characteristic scale length $r_0=4.61^{+1.00}_{-0.68}\, {h^{-1}{\rm Mpc}}$ and slope $\gamma=1.59^{+0.23}_{-0.25}\, {h^{-1}{\rm Mpc}}$. At $5 < z < 6$ it follows $r_0=6.23^{+1.68}_{-1.13}\, {h^{-1}{\rm Mpc}}$ and $\gamma=1.63^{+0.24}_{-0.26}$. When protocluster galaxies are included in the clustering analysis, $\gamma$ gets flattened with larger $r_0$.  The clustering of all HAEs, including protocluster members, shows $r_0=8.29^{+1.66}_{-1.21}\, {h^{-1}{\rm Mpc}}$ and $\gamma=1.34^{+0.13}_{-0.14}$ at $4 < z < 5$, and $r_0=19.71^{+3.38}_{-2.90}\, {h^{-1}{\rm Mpc}}$ and $\gamma=1.02^{+0.03}_{-0.02}$ at $5 < z < 6$.  
    
    \item The flattened slopes of the auto-correlation functions that include protocluster galaxies reflect the geometry of the filamentary structures. The galaxy pairs within filaments lead to elongated patterns in the 2D auto-correlation results. This suggests that galaxy clustering in the presence of protoclusters may not necessarily follow a power-law form with typical slopes of $1.6-2.0$ as reported in the literature, especially in the high-redshift Universe when overdensities and filamentary structures are ubiquitous. 

    \item We compare the measured HAE clustering with \textsc{UniverseMachine} simulations, applying the same selection function. We select regions in \textsc{UniverseMachine} mock surveys where field galaxies or protoclusters dominate star formation to reproduce the observed sample variance. Our measured HAE auto-correlation functions are very consistent with \textsc{UniverseMachine} predictions in both protocluster- and field galaxy-dominated scenarios. \xj{We confirm that filamentary structures drive the differences in clustering in the protocluster-dominated scenario compared to the field, primarily resulting in elevated amplitudes at $r>1~h^{-1}{\rm Mpc}$ and flattened slopes in the auto-correlation functions.} We infer that the masses of host dark matter (sub-)halo are $\log (M_h/M_\odot)=11.21^{+0.36}_{-0.33}$ for HAEs at $4 < z < 5$, and $\log (M_h/M_\odot)=11.06^{+0.36}_{-0.32}$ for HAEs at $5 < z < 6$.
    
\end{itemize}

By studying the \ha\ LFs in protoclusters and fields, we highlight the impact of large-scale environments on galaxy evolution. By performing the first SFR-limited three-dimensional clustering analysis of HAEs at $ z > 4 $, we demonstrate the power of JWST NIRCam/WFSS observations in tracing structure formation and dark matter halos. Future grism surveys with greater depth and larger volumes, such as SAPPHIRES \citep[GO-6434, PI Egami;][]{sapphires_edr} and COSMOS-3D (GO-5893, PI Kakiichi), are crucial for more accurately constraining the cosmic star formation history and clustering properties, mitigating cosmic variance, and \xj{uncovering clustering's dependence on luminosity, stellar mass, and SFR, etc., to connect galaxy properties to dark matter halo properties (halo mass, etc.)}
 
\bigskip

\section*{Data Availability}
The JWST data presented in this article were obtained from the Mikulski Archive for Space Telescopes (MAST) at the Space Telescope Science Institute. The data of the FRESCO survey \citep{FRESCO_hlsp} is available at  \dataset[DOI:10.17909/gdyc-7g80]{https://doi.org/10.17909/gdyc-7g80}; the data of the CONGRESS survey is available at  \dataset[DOI:10.17909/6rfk-6s81]{https://doi.org/10.17909/6rfk-6s81}; the data of the JADES survey \citep{JADES_hlsp} is available at \dataset[DOI:10.17909/8tdj-8n28]{https://doi.org/10.17909/8tdj-8n28}.

\section*{Acknowledgments}

We thank the anonymous referee for providing constructive comments. We thank Nickolas Kokron, Michael A. Strauss, and Yin Li for very helpful discussions on the clustering analysis. X.L. and X.F. acknowledge support from the NSF award AST-2308258. F.W. acknowledges support from  NSF award AST-2513040. X.L. and Z.C. acknowledge support from the National Key R\&D Program of China (grant no. 2023YFA1605600) and Tsinghua University Initiative Scientific Research Program (No. 20223080023). AJB acknowledges funding from the “FirstGalaxies” Advanced Grant from the European Research Council (ERC) under the European Union’s Horizon 2020 research and innovation program (Grant agreement No. 789056). BER acknowledges support from the NIRCam Science Team contract to the University of Arizona, NAS5-02015, and JWST Program 3215. C.N.A.W acknowledges JWST/NIRCam contract to the University of Arizona NAS5-02015. ST acknowledges support by the Royal Society Research Grant G125142.

This work is based on observations made with the NASA/ESA Hubble Space Telescope and NASA/ESA/CSA James Webb Space Telescope. The data were obtained from the Mikulski Archive for Space Telescopes at the Space Telescope Science Institute, which is operated by the Association of Universities for Research in Astronomy, Inc., under NASA contract NAS 5-03127 for JWST. These observations are associated with program \#1181 (JADES), \#1895 (FRESCO), and \#3577 (CONGRESS). Support for program \#3577 was provided by NASA through a grant from the Space Telescope Science Institute, which is operated by the Association of Universities for Research in Astronomy, Inc., under NASA contract NAS 5-03127.
The authors acknowledge the FRESCO team for developing their observing program with a zero-exclusive-access period.

\appendix
\counterwithin{figure}{section}

\section{Emitter catalog and redshift determination algorithm}\label{sec:redshift_algorithm}

In \S\ref{sec:emission_line_selection} we briefly summarized the emission line detection and redshift determination algorithms for JWST/NIRCam WFSS data. Here we provide a more detailed description.

\subsection{Identification of robust emission lines}\label{sec:effective_line_table}

As described in \S\ref{sec:emission_line_selection}, we detect emission lines on the 1D and 2D spectra following \cite{Wang2023}. Then we define a detected emission line as robust when it satisfies all the criteria below.
\begin{itemize}
    \item[(1)] Its line flux cannot exceed the value derived from the photometry in the same filter as the grism spectra.

    \item[(2)] It should be detected simultaneously in both 1D and 2D coadded spectra, and the centers of 2D emission lines should be within $\pm 3$ pixels of the predicted spectral traces.

    \item[(3)] If both module-A and module-B observations are available and cover the position of the emission line, the line should be detected simultaneously in the 1D and 2D spectra of module-A-only and module-B-only extractions.
\end{itemize}
 Criteria (1) and (2) are used to remove contamination from bright sources, especially in cases where the spectra of high-redshift galaxies overlap with those of large, bright, low-redshift objects. For Criterion (3), we predict the line S/N in the single-module extracted spectra according to the line flux measured from the coadded spectra and the exposure times for each module. If the predicted line S/N $<3$, we do not require Criterion (3).  We finally obtain a table of robust emission lines for each source, and only these robust emission lines are used to determine the redshift in the following steps.

 \subsection{Two-step cross-correlation to determine the redshift of emitters}\label{sec:twostep_correlation}

\begin{table*}
    \centering
    \begin{tabular}{c|c}
    \hline
        line basis & line name \\ \hline
        $\eta_1$ & Pa$-\alpha$ + Pa$-\beta$ (+ [\ion{Fe}{2}] $\lambda\lambda$1.644$\mu$m, 1.257$\mu$m) + Pa$-\gamma$ + Pa$-\delta$ \\ \hline
        $\eta_2$ & Br$-\alpha$ + Br$-\beta$ + Br$-\gamma$ + Br$-\delta$ \\ \hline
        $\eta_3$ & \ion{He}{1} + Pa$-\gamma$ \\ \hline
        $\eta_4$ & [\ion{S}{3}] $\lambda\lambda$9069, 9531 \\ \hline
        $\eta_5$ & H$\beta$ + [\ion{O}{3}] $\lambda\lambda$4960,5008 \\ \hline
        $\eta_6$ & \ha\ +  
        [\ion{N}{2}]$\lambda\lambda$6549,6585 + [\ion{S}{2}]$\lambda\lambda$6718,6732 \\ \hline
    \end{tabular}
    \caption{The emission line templates used for the redshift estimation. In this study, the [\ion{Fe}{2}] $\lambda\lambda$1.644$\mu$m and 1.257$\mu$m lines are applied only when the sources are brighter than 21 mag in either the F356W or F444W bands. For each line basis, the emission line ratios are fixed based on literature observations.}
    \label{tab:line_basis}
\end{table*}

We run a two-step cross-correlation algorithm between the robust emission lines and a series of emission-line templates for each source. The procedures are described below.

\noindent\textbf{Step 1: line position cross-correlation} Based on the detected valid emission line table, we construct a model line spectrum, where each line has uniform line flux and FWHMs of the spectral resolution.  Then we generate an emission line template at $z=0-10$ with a step of $\delta z=0.1$, including all emission lines in Table \ref{tab:line_basis}. The wavelength grids of the line templates are matched to those of the observed grism spectra, with all lines assigned uniform fluxes and FWHMs equal to the spectral resolution. At each $z$, we calculate the correlation coefficient between the model line spectrum and the line template.  The resulting correlation coefficients vary as a function of $z$ and will peak at $z$ values where the template emission lines align with those in the observed spectra. In this step, only the wavelengths of the detected emission lines are used; the measured line fluxes are not considered.

\noindent\textbf{Step 2: line spectrum cross-correlation} In the step above, we obtain several redshifts where the cross-correlation has high values. These redshifts are all possible redshift solutions. For each possible redshift, we set finer grids with $\delta z=0.001$ across $z\pm 0.2$.  We set a series of line template bases $\bm{\eta}$ (Table \ref{tab:line_basis}), and each basis $\eta_i$  has fixed line flux ratios. 

\begin{equation}
    \bm{\eta} = (\vec{\eta}_1, \vec{\eta}_2, ..., \vec{\eta}_m)
\end{equation}

We fit $\bm{\eta}$ to the continuum-removed spectrum $\vec{\psi}$. 

\begin{equation}
    \chi^2 = \frac{1}{2} \left(\bm{\eta} \vec{\alpha} - \vec{\psi}\right)^T \cdot \left(\bm{\eta} \vec{\alpha} - \vec{\psi}\right)
\end{equation}
where $\vec{\alpha}=(\alpha_1, ..., \alpha_m)^T$ is the scaling factor of each basis. The best-fit spectrum would be 

\begin{equation}
    \hat{\psi} = \bm{\eta} \vec{\alpha}
\end{equation}

We will finally obtain a series of $\chi^2$ as a function of redshift. The probability of spec-$z$ distribution would be $\ln(z_{\rm spec}) = - \chi(z_{\rm spec})^2$

\subsection{The $z_{\rm best}$ solution and  the confidence level}\label{sec:score_system}
\begin{figure*}
    \includegraphics[width=\textwidth]{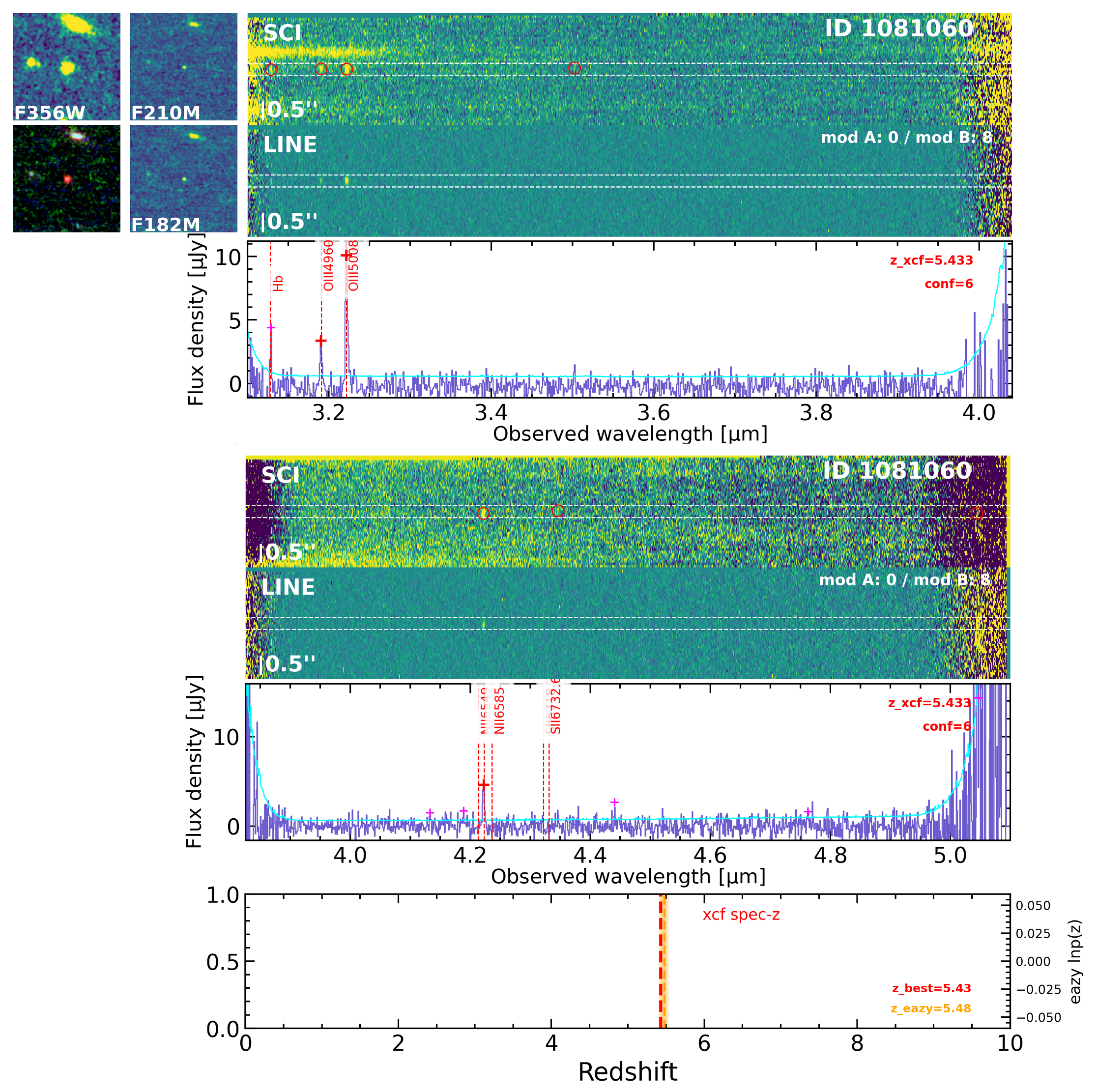}
    \caption{An example of \ha\ emitter at $z=5.4$ with \texttt{Confidence 6}. The top three panels show the original 2D spectrum, the continuum-removed (line-only) spectrum, and the extracted 1D line spectrum of the F356W grism. The middle three panels show the spectra of the F444W grism. In the 2D spectra, the white dashed lines indicate the aperture of 5 pixels ($0\farcs315$), and the red circles mark the detected 2D emission lines. In the 1D spectra of F356W and F444W, the cyan lines represent the errors, the red pluses mark detected emission lines with S/N $> 5$, and the magenta pluses indicate detected lines with S/N $< 5$. In the bottom panel, the phot-$z$ ranges are labeled as an orange-shaded region, with the best phot-$z$ (\texttt{z\_eazy}) indicated by the orange dashed lines. The red dashed line represents the estimated spectroscopic redshift $z_{\rm best}$ (\texttt{z\_xcf}). For this object, three lines with S/N $> 5$ are used by the cross-correlation algorithm to determine the redshift: the [\ion{O}{3}] doublets in the F356W grism and H$\alpha$ in the F444W grism. The resulting $z_{\rm best}$ differs by only 0.05 from the \textsc{eazy}-estimated phot-$z$, so we classify the redshift result as \texttt{Confidence 6}. The positions of H$\beta$, [\ion{O}{3}]$\lambda\lambda$ 4960, 5008, H$\alpha$, [\ion{N}{2}]$\lambda\lambda$6549, 6585, and [\ion{S}{2}]$\lambda\lambda$6718, 6732 at $z_{\rm best}$ are labeled as red dashed lines in the extracted 1D spectra. }
    \label{fig:example_zxcf}
\end{figure*}

\begin{figure}
    \centering
    \includegraphics[width=\linewidth]{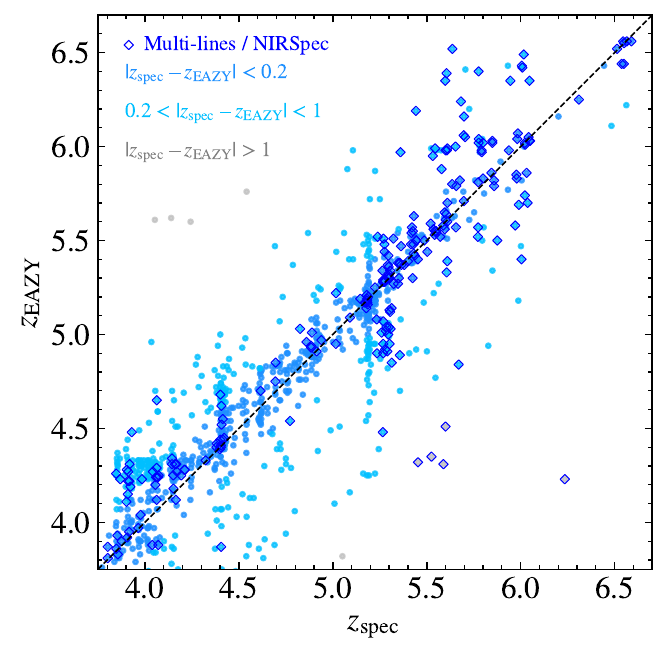}
    \caption{The comparison between photometric redshifts ($z_{\rm EAZY}$) and grism-determined spectroscopic redshifts ($z_{\rm spec}$) for HAEs at $3.75<z<6.6$. The blue, light-blue, and silver dots represent targets with different offsets between $z_{\rm EAZY}$ and $z_{\rm spec}$. For sources with multiple lines detected in either the grism spectra or NIRSpec spectra (i.e., \ha+H$\beta$+[\ion{O}{3}] for emitters at $z>5$), we highlight them using blue diamonds.}
    \label{fig:photz_specz}
\end{figure}

We develop a score system to select the most likely redshift solution and evaluate its confidence level. We first select the top five redshift solutions with the lowest $\chi^2$ values. We then define the best redshift solution, \( z_{\rm best} \), as either the solution among the top five with the most lines used for identification or the solution closest to the photometric redshift, \( z_{\rm phot} \). We assess the following six confidence levels:
\begin{itemize}
    \item \texttt{Confidence 6}: $|z_{\rm best} - z_{\rm phot}|\leq 0.2$ and  $z_{\rm best}$ relies on more than two spectral lines with S/N $>$ 5.
    \item  \texttt{Confidence 5}:  $0.2<|z_{\rm best} - z_{\rm phot}|\leq1$ and $z_{\rm best}$ relies on more than two spectral lines with S/N $>$ 5.
    \item  \texttt{Confidence 4}: $|z_{\rm best} - z_{\rm phot}|>1$ and $z_{\rm best}$ relies on more than two spectral lines with S/N $>$ 5.
    \item   \texttt{Confidence 3}: $|z_{\rm best} - z_{\rm phot}|\leq0.2$, but only one line with S/N $>$ 5 is used at this redshift.
    \item   \texttt{Confidence 2}: $0.2<|z_{\rm best} - z_{\rm phot}|\leq1$ while only one  line with S/N $>$ 5 is used.
     \item   \texttt{Confidence 1}: $|z_{\rm best} - z_{\rm phot}|>1$ while only one  line with S/N $>$ 5 is used.
\end{itemize}
Figure \ref{fig:example_zxcf} presents an example of an H$\alpha$ emitter at $z = 5.4$ with \texttt{Confidence 6}. Its $z_{\rm best}$ is determined by the [\ion{O}{3}] doublets in the F356W and the H$\alpha$ line in the F444W, differing by only 0.05 from $z_{\rm phot}$. We present the statistical comparison between $z_{\rm phot}$ and $z_{\rm spec}$ for the HAEs selected in this work in Figure \ref{fig:photz_specz}. The median $|z_{\rm phot}-z_{\rm spec}|$ is 0.14 and the median $|z_{\rm phot}-z_{\rm spec}|/z_{\rm spec}$ is 0.03.

\section{completeness function}\label{sec:completeness_function}

\begin{figure*}
    \centering
    \includegraphics[width=0.63\linewidth]{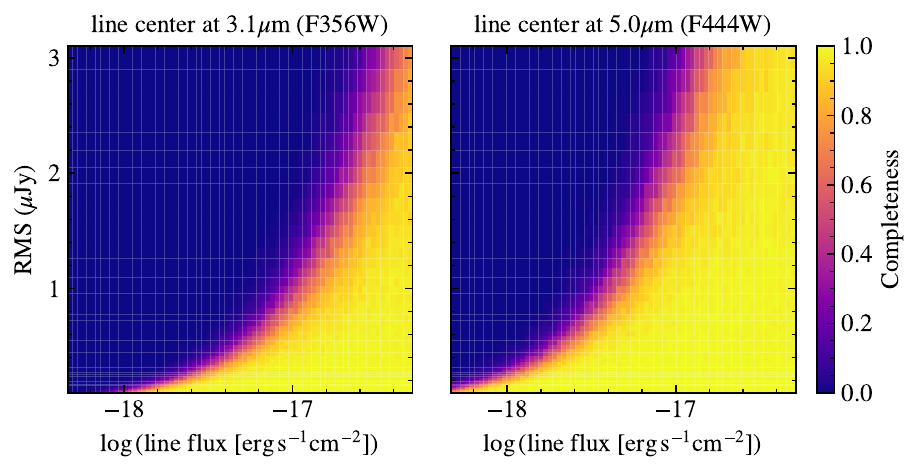}
    \includegraphics[width=0.35\linewidth]{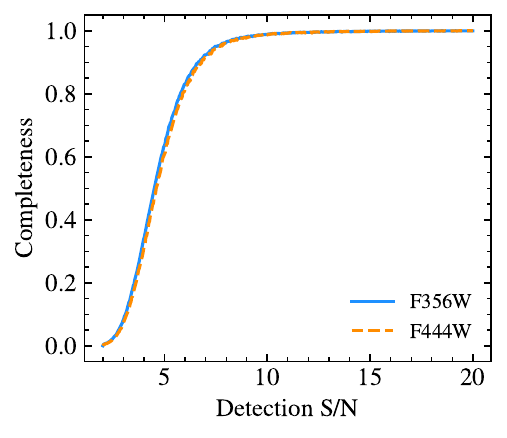}
    \caption{The completeness model used for \ha\ LFs and random galaxy generation. \textit{Left}: Examples of single line completeness at 3.1$\mu$m and 5.0$\mu$m. The completeness is a function of RMS values and line fluxes. \textit{Right}: The median completeness as a function of line S/N during line detection procedure for F356W and F444W.
    }
    \label{fig:completeness}
\end{figure*}

To optimize the efficiency of LF calculations and random catalog generation, we develop a completeness model based on line flux and the RMS values of coadded 1D grism spectra. This approach can be generalized for completeness corrections for any grism-selected line emitters. First, we generate a series of grids for line flux and RMS values at different wavelengths. At a fixed line flux and line wavelength bin, we extract the distribution of \ha\ line FWHM for the observed HAEs. This distribution incorporates both the intrinsic line widths and the broadening due to source morphology. We then generate 1,000 mock line spectra by assuming a Gaussian profile, with the given line flux and FWHM randomly drawn from the extracted distribution above. Then, we add noise to the mock line spectra for each RMS value in the RMS grid, resulting in 1,000 simulated spectra for each combination of line flux and RMS value. We run the same line detection algorithm in these simulated spectra as was done for the real grism data. The fraction of successfully detected mock lines at each flux-RMS combination defines the completeness. The experiments are repeated for emitters with lines centered at various wavelengths. Finally, we obtain a three-dimensional completeness cube as a function of line flux, RMS, and line wavelength.  The left two panels of Figure \ref{fig:completeness} present two example slices of the completeness cube. For any emission line, its completeness can be computed by interpolating the completeness cube using its wavelength, flux, and RMS at the line position. The right panel of Figure \ref{fig:completeness} illustrates how the line completeness varies with the line S/N.

\section{\ha\ luminosity functions}
\subsection{Results of the direct $1/V_{\rm max}$ method.}
The direct 1/$V_{\rm max}$ measurements of \ha\ LFs corresponding to Figure \ref{fig:HaLF} are listed in Table \ref{tab:HaLF}, including the measured number density ($\Phi$), the number of galaxies used ($N_{\rm gal}$) and the average completeness correction ($\langle c \rangle$).
\begin{table*}
	\begin{center}
		\begin{tabular}{c|ccc|ccc}
			\hline
            \hline
			$\log L_{\rm H\alpha}$ & $\Phi$ & $N_{\rm gal}$ & $\langle c \rangle$ & $\Phi$ & $N_{\rm gal}$ & $\langle c \rangle$ \\
			(erg s$^{-1}$) & ($10^{-3}$ Mpc$^{-3}$ dex$^{-1}$) &  &  & ($10^{-3}$ Mpc$^{-3}$ dex$^{-1}$) &  &  \\
            \hline
			& \multicolumn{3}{c|}{All HAEs $3.75<z<5.00$} & \multicolumn{3}{c}{All HAEs $5<z<6$}    \\
         
			41.125 & $0.984 \pm 0.730$ & 2 & 0.042 & $14.549 \pm 14.985$ & 1 & 0.002 \\
			41.375 & $1.849 \pm 0.398$ & 26 & 0.285 & $19.040 \pm 6.160$ & 19 & 0.037 \\
			41.625 & $3.002 \pm 0.258$ & 114 & 0.777 & $4.191 \pm 0.615$ & 50 & 0.381 \\
			41.875 & $3.573 \pm 0.226$ & 174 & 0.948 & $2.840 \pm 0.254$ & 88 & 0.794 \\
			42.125 & $2.456 \pm 0.193$ & 125 & 0.980 & $1.917 \pm 0.196$ & 74 & 0.980 \\
			42.375 & $1.360 \pm 0.150$ & 70 & 0.990 & $1.397 \pm 0.174$ & 56 & 0.994 \\
			42.625 & $0.581 \pm 0.102$ & 30 & 0.993 & $0.583 \pm 0.121$ & 24 & 0.996 \\
			42.875 & $0.329 \pm 0.080$ & 17 & 0.995 & $0.254 \pm 0.076$ & 10 & 0.995 \\
			43.125 & $0.077 \pm 0.037$ & 4 & 1 & $0.076 \pm 0.043$ & 3 & 0.992 \\
			43.375 & $0.019 \pm 0.019$ & 1 & 1 & -- & -- & -- \\
            \hline
		 & \multicolumn{3}{c|}{$3.75<z<5.00$ field} &  \multicolumn{3}{c}{$5<z<6$ field} \\
    
			41.125 & $0.379 \pm 0.371$ & 1 & 0.053 & $18.823 \pm 19.001$ & 1 & 0.002 \\
			41.375 & $1.664 \pm 0.370$ & 23 & 0.292 & $13.351 \pm 7.020$ & 7 & 0.029 \\
			41.625 & $2.529 \pm 0.230$ & 91 & 0.783 & $2.320 \pm 0.489$ & 24 & 0.378 \\
			41.875 & $3.114 \pm 0.210$ & 145 & 0.947 & $1.728 \pm 0.216$ & 44 & 0.773 \\
			42.125 & $2.094 \pm 0.180$ & 102 & 0.979 & $1.021 \pm 0.150$ & 34 & 0.972 \\
			42.375 & $1.056 \pm 0.137$ & 52 & 0.990 & $0.646 \pm 0.128$ & 22 & 0.994 \\
			42.625 & $0.526 \pm 0.099$ & 26 & 0.993 & $0.205 \pm 0.075$ & 7 & 0.996 \\
			42.875 & $0.262 \pm 0.072$ & 13 & 0.995 & $0.088 \pm 0.050$ & 3 & 0.995 \\
			43.125 & $0.020 \pm 0.019$ & 1 & 1 & $0.029 \pm 0.027$ & 1 & 0.993 \\
			43.375 & $0.020 \pm 0.020$ & 1 & 1 & -- & -- & -- \\
            \hline
			& \multicolumn{3}{c|}{$4.39<z<4.45$ protocluster} &  \multicolumn{3}{c}{$5.16<z<5.30$ protocluster}  \\
     
			41.125 & $10.114 \pm 10.487$ & 1 & 0.039 & -- & -- & -- \\
			41.375 & $4.646 \pm 2.636$ & 3 & 0.248 & $35.263 \pm 11.406$ & 12 & 0.086 \\
			41.625 & $11.216 \pm 1.995$ & 23 & 0.816 & $11.336 \pm 2.162$ & 26 & 0.498 \\
			41.875 & $11.817 \pm 1.853$ & 29 & 0.975 & $8.645 \pm 1.144$ & 44 & 0.893 \\
			42.125 & $9.113 \pm 1.733$ & 23 & 0.992 & $7.155 \pm 1.027$ & 40 & 0.988 \\
			42.375 & $7.097 \pm 1.503$ & 18 & 0.997 & $5.843 \pm 0.909$ & 34 & 0.996 \\
			42.625 & $1.579 \pm 0.798$ & 4 & 0.996 & $2.833 \pm 0.691$ & 17 & 0.997 \\
			42.875 & $1.579 \pm 0.777$ & 4 & 0.996 & $1.240 \pm 0.460$ & 7 & 0.996 \\
			43.125 & $1.179 \pm 0.670$ & 3 & 1 & $0.355 \pm 0.247$ & 2 & 0.993 \\
			\hline
		\end{tabular}
    \caption{The observed \ha\ LF at $3.75 < z < 5$ and $5 < z < 6$ measured using the direct $1/V_{\rm max}$ method. In the \textsc{All HAEs} scenario, we include all HAEs and their associated survey volumes within the FoV. In the \textsc{field} scenario, we exclude the survey volume dominated by protoclusters. In the \textsc{protocluster} scenario, we consider only the protocluster member galaxies and the corresponding survey volume they occupy, as defined by the redshift range. $\Phi$ represents the completeness-corrected galaxy number density in each luminosity bin, $N_{\rm gal}$ is the number of galaxies used, and $\langle c \rangle$ is the median completeness value in each luminosity bin.}
    \label{tab:HaLF}
	\end{center}
\end{table*}

\subsection{Results of the MLE method}\label{sec:Appendix_MLE_result}

\begin{figure*}
    \centering
    \includegraphics[width=0.8\linewidth]{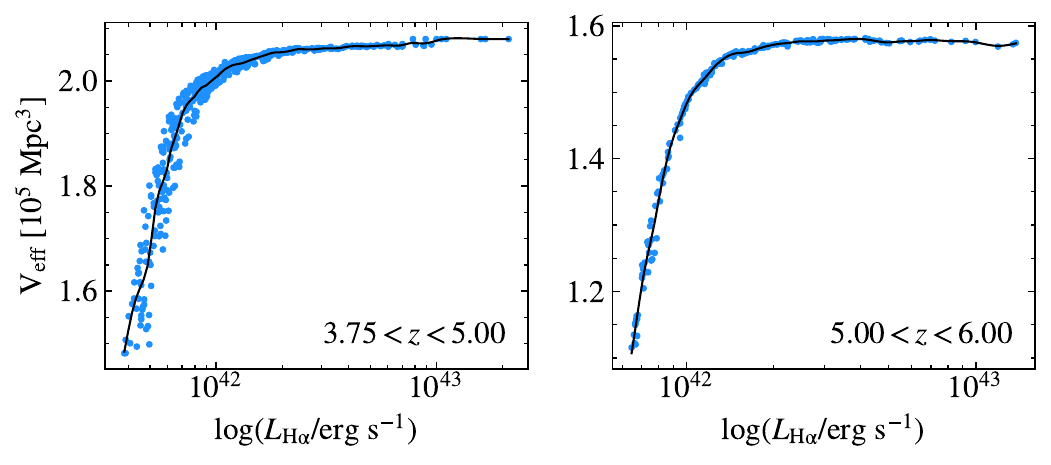}
    \caption{The $V_{\rm eff}$ as a function of H$\alpha$ luminosity. The blue dots denote the $V_{{\rm eff},i}$ of individual galaxies, i.e., $C_i V_{{\rm max},i}$, and the black line represents a spline function fit to the $V_{\rm eff}$. }
    \label{fig:Veff_spline}
\end{figure*}

In \S\ref{sec:HaLF}, we also use the MLE method to model the Schechter function of H$\alpha$ LFs. In Equation \ref{eq:MLE_P}, we adopt $C_i V_{{\rm max},i}$ as the effective volume for individual galaxies, $V_{{\rm eff},i}$. To obtain $V_{{\rm eff}}(L)$ as a function of H$\alpha$ luminosity, we fit a spline function to the $V_{{\rm eff},i}$, as shown in Figure \ref{fig:Veff_spline}. The average $V_{{\rm eff}}$ can be well described by a smooth curve as a function of H$\alpha$ luminosity.

\begin{figure*}
    \includegraphics[width=\textwidth]{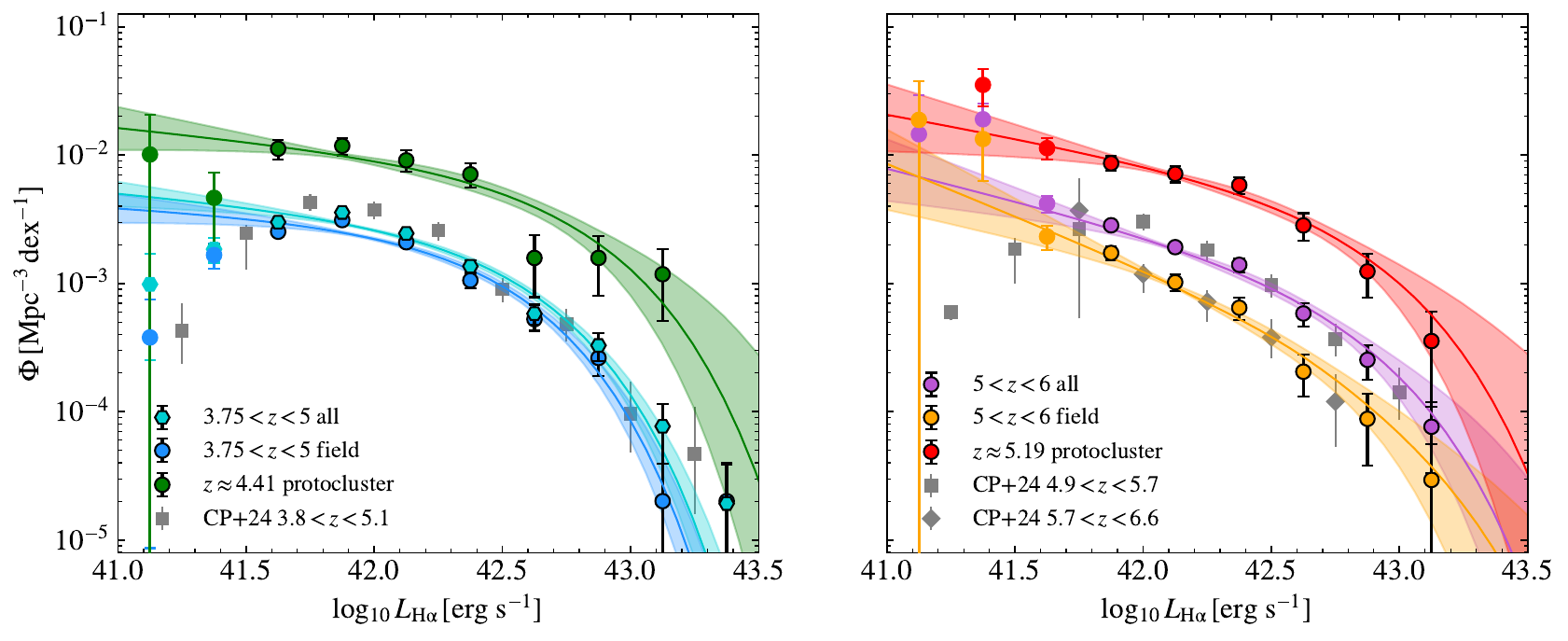}
    \caption{Similar to Figure \ref{fig:HaLF}, but the best-fit models and the corresponding uncertainties are derived from the MLE method.}
    \label{fig:half_mle}
\end{figure*}

\begin{figure}
    \centering
    \includegraphics[width=\linewidth]{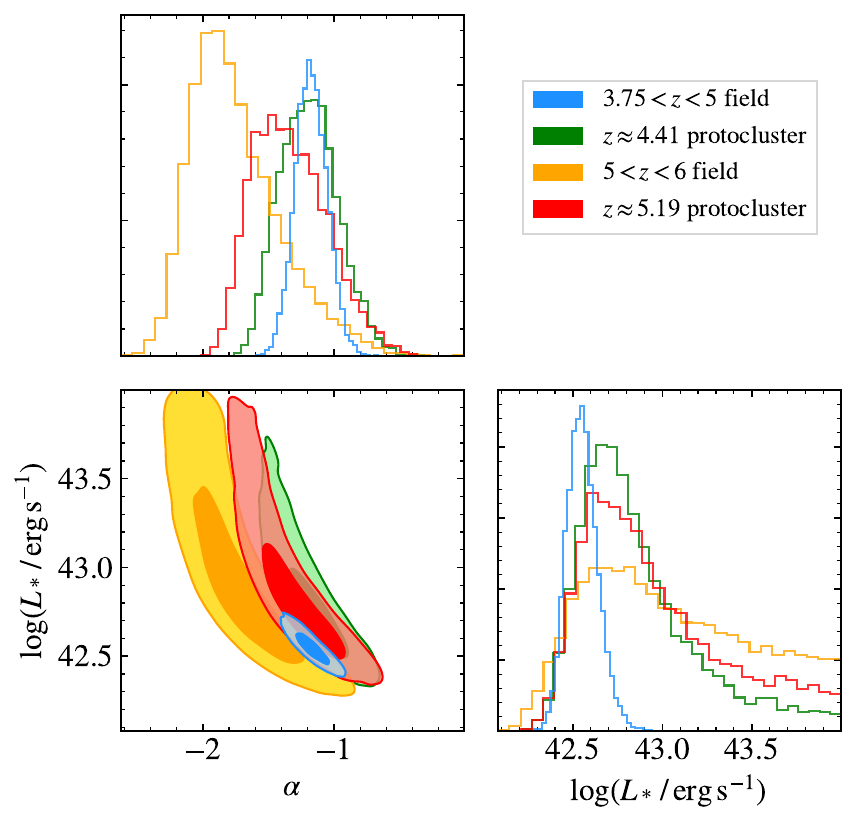}
    \caption{Similar to Figure \ref{fig:HaLF_param_prob}, but the posterior probability distribution of $\log L_*$ and $\alpha$ derived from the MLE method.}
    \label{fig:HaLF_param_prob_mle}
\end{figure}

We show the best-fit models of the H$\alpha$ LFs derived from the MLE method in Figure \ref{fig:half_mle}. The posterior probability distributions of $\log L_*$ and $\alpha$ derived from the MLE method are presented in Figure \ref{fig:HaLF_param_prob_mle}. The $\alpha$ values for the $3.75 < z < 5$ field LF, the $z \approx 4.41$ protocluster, and the $z \approx 5.19$ protocluster LF are consistent with each other within 1$\sigma$. In contrast, the $\alpha$ for the $5 < z < 6$ field LF is steeper, reaching $-1.8$. This result is consistent with that derived from the direct $1/V_{\rm max}$ method, validating the conclusion that protocluster galaxies at $z > 5$ are more evolved than field galaxies and have already established LF shapes.

\subsection{Results of Lynden-Bell's $C^-$ method}\label{sec:appendix_lf_cm}

We further estimate the LFs using Lynden-Bell's $C^-$ method \citep{Lynden-Bell1971, Woodroofe1985, Wang1986}, which has previously been applied to quasar LFs \citep[e.g.,][]{Fan2001}. The $C^-$ method is an unbinned, non-parametric maximum likelihood estimator of the cumulative luminosity function. It properly accounts for truncation effects, i.e., the boundaries imposed by the selection function on the luminosity–redshift plane in a flux-limited survey, and is statistically robust against binning effects.

\begin{figure*}[htbp]
    \includegraphics[width=\textwidth]{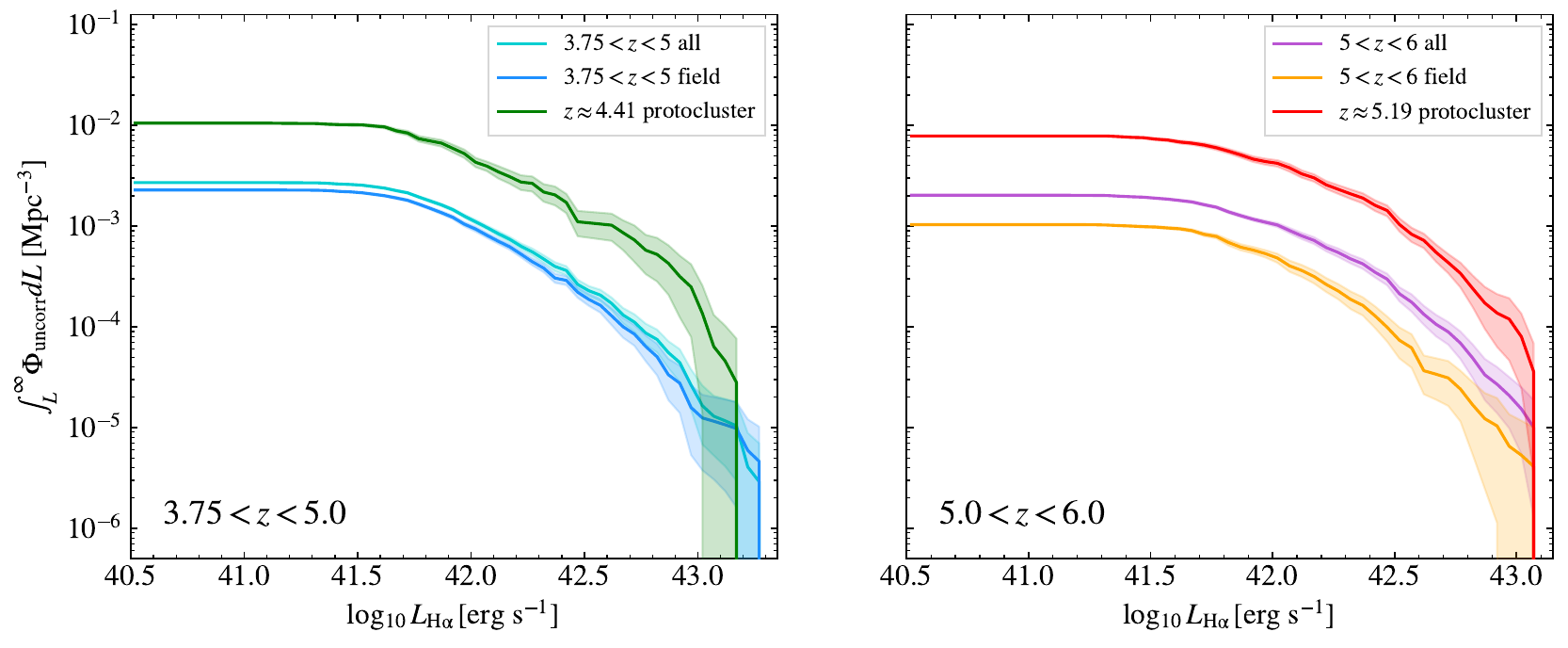}
    \caption{The cumulative LFs, uncorrected for completeness, measured using Lynden-Bell's $C^-$ method.}
    \label{fig:cumulative_LF}
\end{figure*}

\begin{figure*}[htbp!]
    \includegraphics[width=\textwidth]{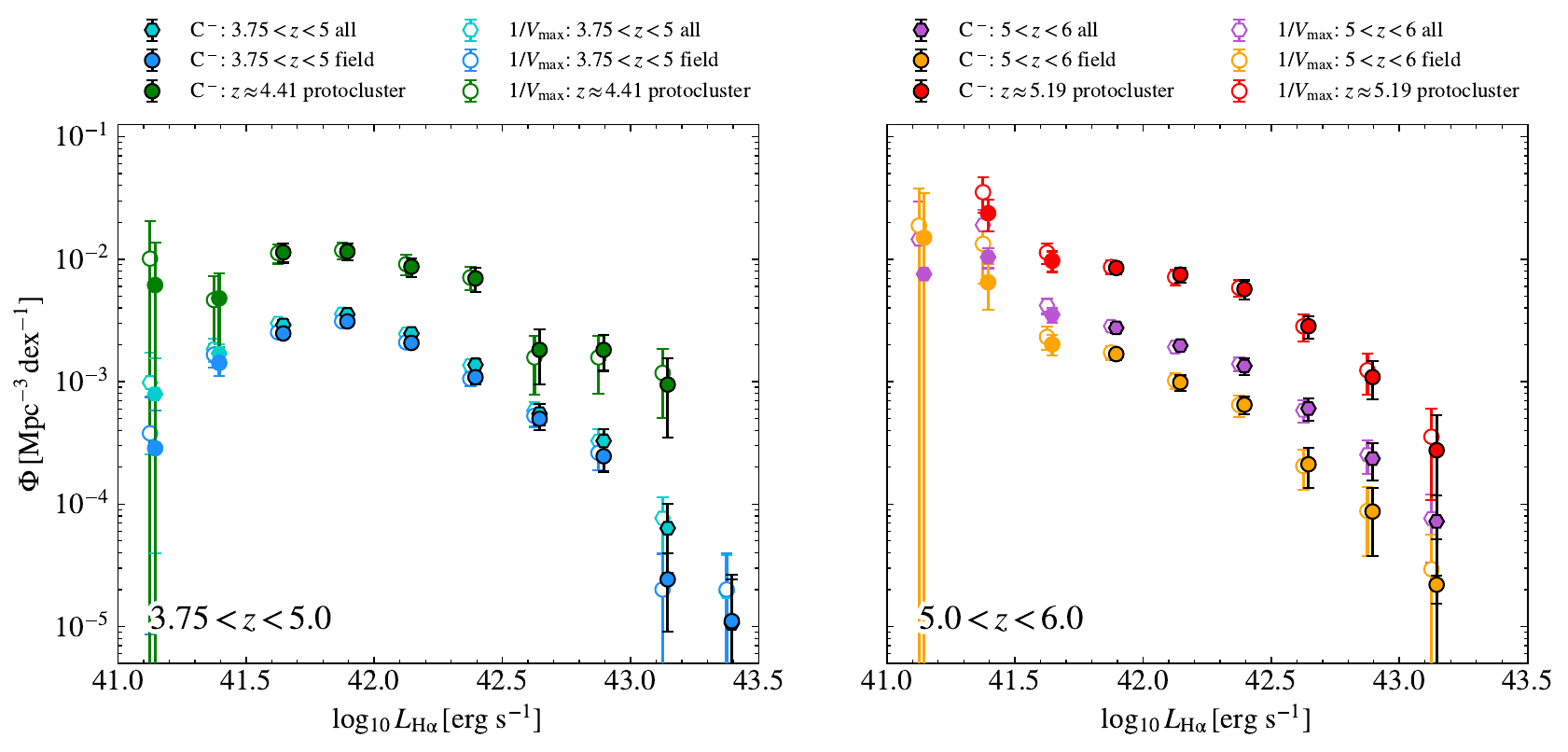}
    \caption{Comparison of LFs measured using 1/$V_{\rm max}$ and Lynden-Bell's $C^-$ method. Filled markers show measurements from the $C^-$ method, while white-filled markers correspond to 1/$V_{\rm max}$. Measurements without black edges indicate bins with an average completeness below 0.7. The $C^-$ luminosity bins ($x$-axis) are offset by 0.02 dex from the 1/$V_{\rm max}$ bins for clarity.}
    \label{fig:LF_cm}
\end{figure*}

We measure the accumulated LFs using the $C^-$ estimator\footnote{\url{https://www.astroml.org/modules/generated/astroML.lumfunc.bootstrap_Cminus.html}}, as shown in Figure \ref{fig:cumulative_LF}. We differentiate the cumulative LFs, bin the results, apply completeness corrections in each luminosity bin, and estimate the uncertainties via bootstrapping. The LFs of the $C^-$ estimator and their comparison to the $1/V_{\rm max}$ results are shown in Figure~\ref{fig:LF_cm}. The $C^-$ results are well consistent with those from $1/V_{\rm max}$, with all bins agreeing within $1\sigma$.

\section{calibrate UniverseMachine simulations}\label{sec:um_calibration}

\begin{figure*}[htbp]
    \includegraphics[width=\textwidth]{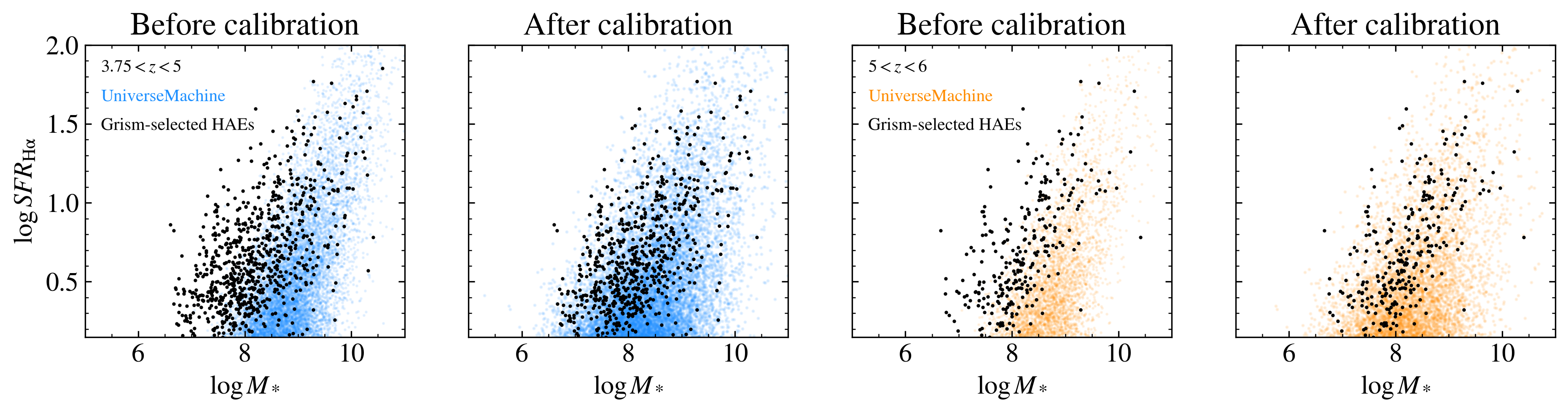}
    \caption{The SFR-$M_*$ distribution of \textsc{UniverseMachine} galaxies before and after the calibration (Appendix \ref{sec:um_calibration}) at $3.75<z<5$ and $5<z<6$. For grism-selected HAEs (black dots), the $M_*$ are from \textsc{beagle} SED fitting, and SFRs are \ha-converted. For \textsc{UniverseMachine} galaxies (blue and orange dots), the $M_*$ is from \textsc{UniverseMachine} lightcone catalogs. In the first and third panels (before calibration), the SFR (y-axis) represents the mock H$\alpha$-corresponding SFRs based on the \textsc{UniverseMachine} total SFRs and dust attenuation. In the second and fourth panels (after calibration), the SFR (y-axis) corresponds to the calibrated value as given by Equation \ref{eq:recalibrated_SFR}, with dust attenuation applied.}
    \label{fig:recalibrated_SFR}
\end{figure*}

As mentioned in \S\ref{sec:um}, the SFRs output by \textsc{UniverseMachine} are calibrated using pre-JWST observational data and do not account for the burstiness of star formation as revealed by recent JWST findings. Therefore, these SFRs cannot be directly converted into \ha\ luminosities. As shown in Figure \ref{fig:recalibrated_SFR}, the \textsc{Universemachine} SFRs are offset from the \ha-converted SFRs of grism-selected HAEs. To derive SFRs that can reasonably reflect \ha\ intensity, we calibrate the \textsc{UniverseMachine} SFRs using their sSFRs, $M_*$, and dust attenuation $A_{\rm UV}$. We assume that the \textsc{UniverseMachine} $M_*$ is accurate, but the sSFRs do not fully capture the bursty star formation and require adjustments.  Galaxies with lower $M_*$ are more likely to exhibit burstier star formation and thus higher sSFR. In this case, the re-calibrated SFR that corresponds to \ha\  can be expressed as
\begin{equation}\label{eq:recalibrated_SFR}
    {\rm SFR_{H\alpha}} = (\mathrm{sSFR_{UM}} + \alpha) \cdot \mathrm{M_{*, UM}} + \beta,
\end{equation}
where $\mathrm{sSFR_{UM}}$ and $\mathrm{M_{*, UM}}$ are the output sSFR and stellar mass in the \textsc{UniverseMachine} lightcone catalogs, respectively. The parameters $\alpha$ and $\beta$ account for the offset between $\mathrm{sSFR_{UM}}$ and $\rm SFR_{ H\alpha}$ due to bursty star formation. We generate a grid of $\alpha$ and $\beta$ combinations. For each $\alpha$ and $\beta$, we convert the ${\rm SFR_{H\alpha}}$ into the intrinsic \ha\ luminosity by 
\begin{equation}
\log \left(L_{\mathrm{H} \alpha} / \mathrm{erg} \mathrm{~s}^{-1}\right)=\log \left(\mathrm{SFR_{\rm H\alpha}} / M_{\odot} \mathrm{yr}^{-1}\right)+41.35.
\end{equation}
To obtain the observed \ha\ luminosity, $L_{\rm H\alpha, obs}$, we then apply dust attenuation using $A_{\rm UV}$ from \textsc{UniverseMachine} at rest-frame 1500 \AA, assuming the SMC dust attenuation law. 

\begin{figure}
    \centering
    \includegraphics[width=\linewidth]{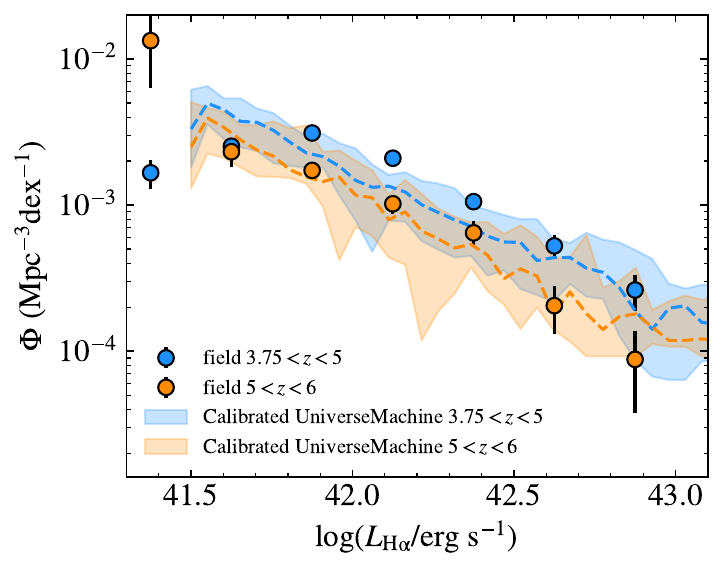}
    \caption{The measured field H$\alpha$ LFs are compared with the calibrated \textsc{UniverseMachine} field H$\alpha$ LFs. The dashed lines represent the median mock field H$\alpha$ LFs. The mock and observed field H$\alpha$ LFs show good agreement within the luminosity range of $10^{41.5}\, {\rm erg\,s^{-1}} < L_{\rm H\alpha} < 10^{43} \, {\rm erg\,s^{-1}}$.}
    \label{fig:UM_calib_HaLF}
\end{figure}

We randomly select ten $32\times 32$ arcmin$^{2}$ \textsc{UniverseMachine} lightcones and place an $8\times 8$ arcmin$^{2}$ aperture in each to construct mock surveys.  To avoid the variance introduced by protoclusters, we run the friends-of-friends algorithm in each mock survey, identify protoclusters that contain more than 30\% of the galaxies within the $8\times 8$ arcmin$^{2}$ region, and exclude the corresponding redshift slices. We then calculate the field \ha\ LFs in each mock surveys and compute the average $\chi^2$ between the mock \ha\ LFs and the observed field \ha\ LFs (\S\ref{sec:HaLF}). As a result,  the average $\chi^{2}$ is a function of $\alpha$ and $\beta$. We limit the H$\alpha$ luminosity range of this comparison to $\log (L_{\rm H\alpha}/{\rm erg\,s^{-1}}) = 41.5 - 43.0$. The lower limit corresponds to the minimum H$\alpha$ luminosity of galaxies included in the clustering analysis. We do not compare the LF at $\log (L_{\rm H\alpha}/{\rm erg\,s^{-1}}) > 43.0$ because, at this range, \textsc{UniverseMachine} predicts a higher number density that cannot be fully addressed by the simple calibration. However, the discrepancy at the very bright end of the H$\alpha$ LF between the simulation and observations does not significantly affect our results, as such galaxies have orders of magnitude lower number densities compared to the rest. 
We find the minimum $\chi^2$ with  
$(\alpha, \beta)=(-10^{-9.70}, 0.4)$ at $z=3.75-5$ and $(\alpha, \beta)=(-10^{-9.70}, 0.2)$ at $z=5-6$.  We compare the calibrated \textsc{UniverseMachine} field \ha\ LFs and the measured field \ha\ LFs in the two redshift bins in Figure \ref{fig:UM_calib_HaLF}.

We further validate the calibration by comparing the \textsc{UniverseMachine} SFR-\( M_* \) relation with the SED-derived values. For each \textsc{UniverseMachine} object, we obtain the observed SFR based on the calibrated SFR$_{\rm H\alpha}$ (Equation \ref{eq:recalibrated_SFR}) with dust attenuation applied.  For each grism-selected HAE, we perform SED modeling using \textsc{beagle}, assuming a constant star formation history along with the SMC dust attenuation law. As shown in Figure \ref{fig:recalibrated_SFR}, the calibrated SFR-$M_*$ relation is in good agreement with the relation between \ha-derived SFRs and SED-derived $M_*$.

As discussed in \S\ref{sec:um_od_num}, we investigate the impact of the calibration parameters $\alpha$ and $\beta$ on the number of protoclusters in \textsc{UniverseMachine}. We select protoclusters from mock surveys calibrated with ($\alpha$, $\beta$) = ($-10^{-9}$, 0), ($-10^{-10}$, 0), ($-10^{-9}$, 0.75), and ($-10^{-10}$, 0.75). These $\alpha-\beta$ combinations correspond to the corners of the $\chi^2$ plane where $\chi^2 < 1.0$ or 1.5 for $z=4-5$ and $z=5-6$. At $z=4-5$, we find 1-14 prominent protoclusters across all \textsc{UniverseMachine} mock surveys with the new $\alpha-\beta$ combinations. The probability of finding such structures spans from 0.05\% to 1.9\%. At $z=5-6$, the number of prominent protoclusters ranges from 1 to 11, with the probability ranging from 0.05\% to 2.4\%. We thus qualitatively conclude that, regardless of the specific choice of $\alpha$ and $\beta$, the likelihood of finding significant protoclusters in \textsc{UniverseMachine} remains low. We note that here, we only discuss protoclusters that are as prominent as those observed in the GOODS-N field. Overdensities or galaxy associations with smaller sizes are beyond the scope of this paper.

\bigskip

\bibliography{main}{}
\bibliographystyle{aasjournal}

\end{document}